\documentclass[traditabstract]{aa}  

\usepackage{ae,aecompl}
\usepackage{amsmath}
\usepackage{amssymb}
\usepackage{booktabs}
\usepackage[T1]{fontenc}
\usepackage{graphicx}
\usepackage{hyperref}
\usepackage{lipsum}
\usepackage{multirow}
\usepackage{natbib}
\usepackage{newtxtext,newtxmath}
\usepackage{soul}
\usepackage{Times}

\newcommand{\add}[1]{{\color{black}#1}}

\newcommand{\bm}[1]{{${#1}$}}

\bibpunct[ ]{(}{)}{;}{a}{}{,}

\hypersetup{
     colorlinks   = True,
     citecolor    = blue,
     linkcolor = blue,
     urlcolor = blue
}

\begin{document} 


\title{VIS$\boldsymbol{^3}$COS: I. survey overview and the role of environment and stellar mass on star formation\thanks{Based on observations obtained with VIMOS on the ESO/VLT under the programmes 086.A-0895, 088.A-0550, and 090.A-0401.}
}

\authorrunning{A. Paulino-Afonso et al.}
\titlerunning{The role of environment and stellar mass on star formation}

\author{Ana Paulino-Afonso\inst{1,2,3}\fnmsep\thanks{E-mail: aafonso@oal.ul.pt}
\and
David Sobral\inst{3,4}
\and
Behnam Darvish\inst{5}
\and
Bruno Ribeiro\inst{4,6}
\and
Andra Stroe\inst{7}\fnmsep\thanks{ESO fellow}
\and
Philip Best\inst{8}
\and
Jos\'e Afonso\inst{1,2}
\and 
Yuichi Matsuda\inst{9}
}

\institute{
Instituto de Astrof\'isica e Ci\^encias do Espa\c{c}o, Universidade de Lisboa, OAL, Tapada da Ajuda, PT1349-018 Lisboa, Portugal
\and
Departamento de F\'isica, Faculdade de Ci\^encias, Universidade de Lisboa, Edif\'icio C8, Campo Grande, PT1749-016 Lisboa, Portugal
\and
Department of Physics, Lancaster University, Lancaster, LA1 4YB, UK
\and
Leiden Observatory, Leiden University, P.O. Box 9513, NL-2300 RA Leiden, The Netherlands
\and
Cahill Center for Astrophysics, California Institute of Technology, 1216 East California Boulevard, Pasadena, CA 91125, USA
\and
Centro de Computa\c{c}\~{a}o Gr\'afica, CVIG,  Campus de Azur\'em, PT4800-058 Guimar\~{a}es, Portugal
\and
European Southern Observatory, Karl-Schwarzschild-Str. 2, 85748, Garching, Germany
\and
Institute for Astronomy, University of Edinburgh, Royal Observatory, Blackford Hill, Edinburgh EH9 3HJ, UK
\and
National Astronomical Observatory of Japan, Osawa 2-21-1, Mitaka, Tokyo 181-8588, Japan
}

\date{ }

\abstract{We present the VIMOS Spectroscopic Survey of a Supercluster in the COSMOS field (VIS$^3$COS) at $z\sim0.84$. We use VIMOS high resolution spectra (GG475 filter) to spectroscopically select 490 galaxies in and around the superstructure and an additional 481 galaxies in the line of sight. We present the redshift distribution, the catalogue to be made public, and the first results on the properties of individual galaxies and stacked spectra \add{($3500$ \AA\ ${<\lambda<4200}$ \AA\ rest-frame)}. We probe a wide range of densities and environments (from low-density field to clusters and rich groups). We find a decrease in the median star formation rate from low to high-density environments in all bins of stellar mass and a sharp rise of the quenched fraction \add{(from ${\sim10}$\% to ${\sim40-60}$\%)} of intermediate stellar mass galaxies {$(10<\log_{10}\left(M_\star/\mathrm{M_\odot}\right)<10.75)$} from filaments to clusters. The quenched fraction for massive galaxies shows little dependence on environment \add{being constant at ${\sim30-40}$\%}. 
\add{We find a break in the strength of the [O{\sc ii}] emission, with nearly constant line equivalent widths at lower densities (${\sim-11}$ \AA) and then a drop to ${\sim-2.5}$ \AA\ towards higher densities. The break in the [O{\sc ii}] line strength happens at similar densities (${\log_{10}(1+\delta)\sim0.0-0.5}$) as the observed rise in the quenched fraction.} Our results may provide further clues regarding the different environmental processes affecting galaxies with different stellar masses and highlight the advantages of a single dataset in the COSMOS field probing a wide range of stellar masses and environments. We hypothesize that quenching mechanisms are enhanced in high density regions.}

\keywords{galaxies: clusters -- galaxies: evolution -- galaxies: high redshift -- large-scale structure of Universe}

   \maketitle



\section{Introduction}\label{section:introduction}

In the local Universe, we observe differences in a wide range of galaxy properties (e.g. colours, star formation, morphology) with respect to the environment they reside in \citep[e.g.][]{oemler1974,dressler1980,dressler1984}. Cluster galaxies are typically red and passive, while in low-density environments the population is dominated by blue star-forming galaxies \citep[e.g.][]{dressler1980,balogh2004,kauffmann2004,baldry2006,bamford2009}. The star formation rate (SFR) and star-forming fraction ($f_\mathrm{SF}$) have also been found to correlate strongly with the projected galaxy density \citep[e.g.][]{lewis2002,gomez2003,hogg2004,best2004,kodama2004,peng2010b,darvish2016,cohen2017}. Observations also imply that the most massive galaxies assembled their stellar mass more quickly and had their bulk of star formation quenched at $z\gtrsim1$ \citep[e.g.][]{iovino2010}. While stellar mass and environmental density correlate, it is now possible to disentangle their roles and show that both are relevant for quenching star formation \citep[e.g.][]{peng2010b,sobral2011,muzzin2012,darvish2016}.

Globally, observations show that the star formation rate density ($\rho_{\rm SFR}$) peaks at $z\sim2-3$ and has been declining ever since \citep[e.g.][]{lilly1996,karim2011,burgarella2013,sobral2013,madau2014,khostovan2015}. However, surprisingly, the decline of $\rho_{\rm SFR}$ with increasing cosmic time is happening in all environments \citep[e.g.][]{cooper2008,koyama2013}. Recent studies have also shed more light on when the dependency of star-forming galaxies on environment start to become observable \citep[e.g.][]{scoville2013,darvish2016}. However, it is still unclear exactly how the environment affected the evolution of galaxies and how that may have changed across time. In order to properly answer such questions it is mandatory to conduct observational surveys at high redshift \citep[e.g.][]{tadaki2012,koyama2013,lemaux2014,cucciati2014,shimakawa2018} which can then be used to test theoretical models of galaxy evolution \citep[e.g.][]{volgersberger2014,genel2014,henriques2015,schaye2015,crain2015}.

There have been a plethora of surveys of clusters and their surroundings at $z\lesssim1$ \citep[e.g.][]{treu2003,cooper2008,poggianti2009,lubin2009,cucciati2010a,iovino2010,li2011,muzzin2012,mok2013,koyama2013,lemaux2014,cucciati2014,cucciati2017} with a key focus on the influence of environment on the star formation of galaxies. \add{Emission line surveys of clusters at lower redshifts (${z\sim0.1-0.5}$) targeting either H$\alpha$ \citep[e.g.][]{balogh2002,stroe2015,stroe2017,sobral2016,delpino2017} or [O{\sc ii}] \citep[e.g.][]{nakata2005} find that star formation is suppressed in cluster environments. This suppression seems to be more effective for early-type galaxies \citep[e.g.][]{balogh2002} and to be a slow acting mechanism that mainly affects the gas component \citep[e.g.][]{delpino2017}.} 

By $z\sim1$, some authors have claimed to have found a flattening, or even a definitive reverse, of the relation between the star formation activity and the projected local density, either studying how the average SFRs of galaxies change with local density \citep[][]{elbaz2007} or looking at $f_\mathrm{SF}$ as a function of density \citep[e.g.][]{ideue2009,tran2010,santos2014}. These results would be naturally interpreted as a sign of evolution if other studies \citep[e.g.][]{patel2009, sobral2011, muzzin2012, santos2013} had not found an opposite result. The differences found between different clusters may be related to their dynamical state, as merging clusters in the low-redshift Universe can also show reverse trends when compared to relaxed clusters at similar epochs \citep[e.g.][]{stroe2014,stroe2015a,stroe2017,mulroy2017}, but other factors like sample size, AGN contamination, environments probed may also play a role \citep[e.g.][]{darvish2016}. \cite{sobral2011}, probing a wide range of environments and stellar masses, were able to recover and reconcile the previous apparently contradictory results. They attribute the discrepancies to selection effects. If one restricts to similar stellar masses and/or densities we can find similar trends in different studies. \cite{sobral2011} also separated the individual roles of mass and environment in galaxy evolution \citep[see also][]{iovino2010,cucciati2010b,peng2010b,li2011}.

Finding the exact mechanisms of galaxy quenching and their physical agents is still one of the unsolved problems in galaxy evolution. Many internal (e.g. stellar and AGN feedback) and external (e.g. galaxy environment) physical drivers are thought to be linked to the quenching process. One might naively expect a continuous decline in the star formation of galaxies from the field to the dense cores of clusters (e.g. due to a lower amount of available gas or faster gas consumption as galaxies move trough denser mediums). However, before galaxies undergo a full quenching process in dense regions, they may experience a temporary enhancement in star formation activity \citep[see e.g.][]{sobral2011} which may complicate how observations are interpreted (e.g. ram pressure stripping - \citealt{gallazzi2009,bekki2009,owers2012,roediger2014} - and/or tidal interactions - \citealt{mihos1996,kewley2006,ellison2008}).

When looking in more detail at galaxies in the low to intermediate redshift Universe ($z\lesssim1$), many properties of star-forming galaxies that are directly or indirectly linked to star formation activity (e.g. SFR, sSFR, emission line equivalent widths and the main-sequence of star-forming galaxies) seem to be invariant to their environment \citep[{but it is still a debated issue, see} e.g.][]{peng2010b,iovino2010,wijesinghe2012,muzzin2012,koyama2013,koyama2014,hayashi2014,darvish2014,darvish2015,darvish2016}. Therefore, the main role of the environment seems to be to set the fraction of quiescent/star-forming galaxies \citep[e.g.][]{peng2010b,cucciati2010b,sobral2011,muzzin2012,darvish2014,darvish2016} which is likely linked to the reported a gas deficit in cluster galaxies \citep[seen in atomic hydrogen, e.g.][]{giovanelli1985,cortese2010,serra2012,brown2017}. But this is not the picture found when looking at molecular hydrogen which is either independent of environment, depressed or enhanced in high density regions dependent on the study \citep[e.g.][]{boselli2014b,mok2016,koyama2017}. Nevertheless, recent studies are finding that not all characteristics of star-forming galaxies are independent of environment. For example, metallicities have been shown to be a function of environment \citep[e.g.][]{kulas2013,shimakawa2015,sobral2015b} with studies finding that star-forming galaxies have slightly higher metallicities in high density environments when compared to lower density/more typical environments at $z\sim0.2-0.5$ \cite[e.g.][]{sobral2015b,darvish2015}. \citet{sobral2015b} study a cluster undergoing a merger and \citet{darvish2015} focus on galaxy filaments which are both regions of enhanced dynamical activity. Denser environments also seem to boost the dust content of star-forming galaxies \citep[e.g.][]{koyama2013,sobral2016}. The higher dust content seen in high density regions can be a requirement for galaxies to sustain star formation in such environments, by allowing for dense and compact regions to survive environmental stripping.

Issues related to photometric redshift errors and projection effects can limit our understanding of what is occurring in and around clusters. These issues dilute genuine trends and prohibit us from unveiling the role of the environment in sufficient detail to really test our understanding. Surveys such as EDisCS \citep[e.g.][]{white2005} have aimed to overcome some of these issues by targeting the densest regions at high redshift with extensive spectroscopic observations. These have made significant progress \citep[][]{poggianti2006,poggianti2009,cucciati2010b,cucciati2017}, but either they target deep and small areas or shallow and wide areas. This limits the study on the role of the larger-scale structure and the densest environments simultaneously. A way to make further progress is to conduct a spectroscopic survey (to avoid projection effects and photometric redshift biases and errors) over a superstructure containing the complete range of environments in a sub-deg$^2$ area at high redshift.

\begin{figure*}
\centering
\includegraphics[width=\linewidth]{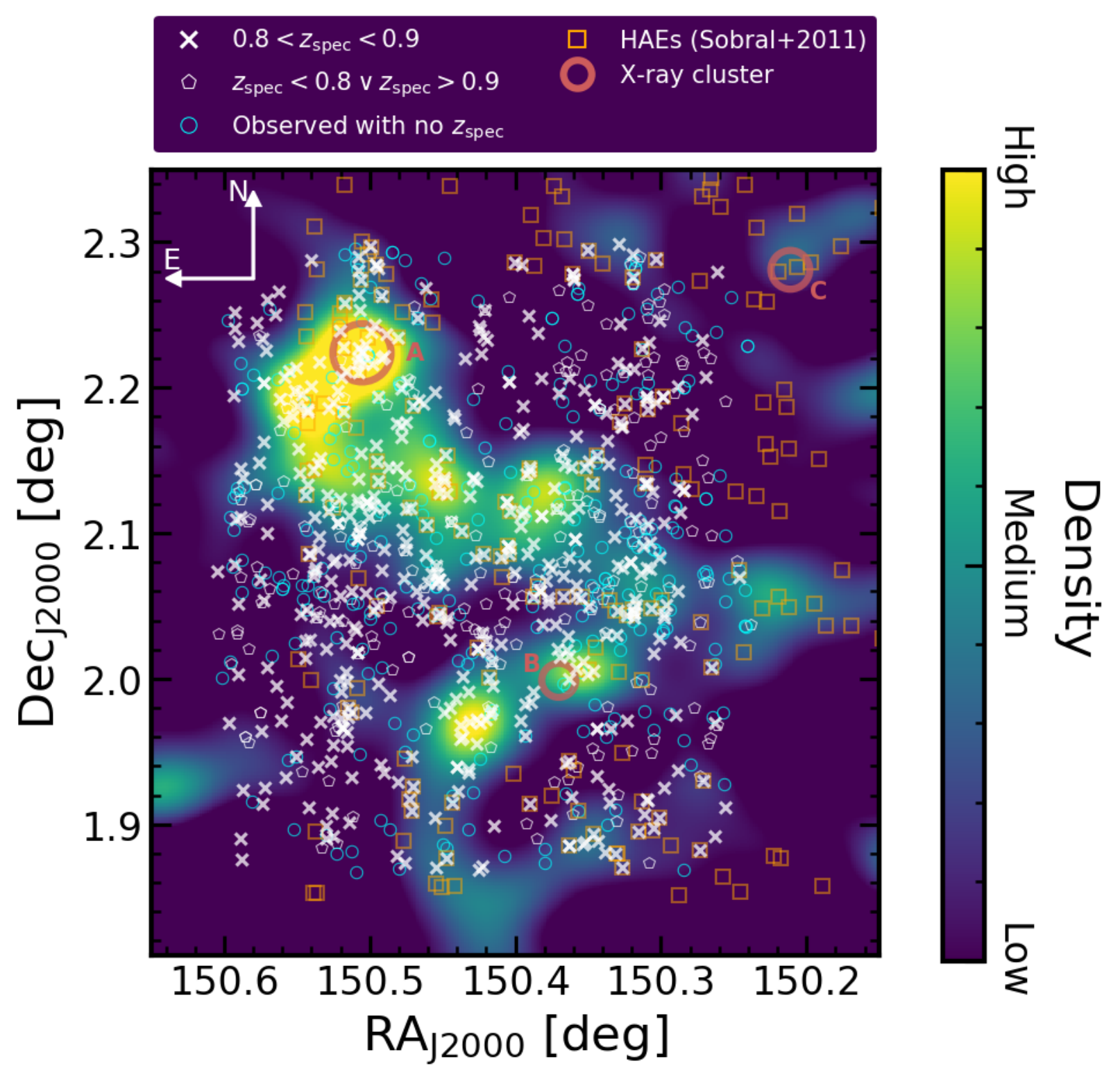}
\caption{Snapshot of the region targeted by our spectroscopic survey. The colormap encodes the information on the galaxy projected surface density at the redshift slice of interest at $0.8<z<0.9$ estimated from the catalogue made public by \citet{darvish2017}. Each thick white cross represents a targeted galaxy with a measured spectroscopic redshift in the same redshift slice. The white pentagons show a targeted galaxy with measured spectroscopic redshift, outside the defined redshift slice. The blue circles show the targeted galaxies for which we have not a measured spectroscopic redshift. The orange squares show the location of H$\alpha$ emitters studied by \citet{sobral2011}. The \add{large red circles} denote the location of X-ray detected clusters from \citet{finoguenov2007} at the same redshifts. The size of the circle  shows the cluster estimated X-ray radius $r_{500}$. We see here that we are probing a large range of densities with our survey, in part due to selection effects (e.g. slit placement constraints).}
\label{fig:cluster_region}
\end{figure*}

In this paper, we present a large spectroscopic follow-up of members of a supercluster in the COSMOS field first detected in X-rays \citep{finoguenov2007} and later in H$\alpha$ {\citep[see Figure \ref{fig:cluster_region},][]{sobral2011}}. We organize this paper as follows. Section \ref{section:data} discusses the sample and presents the observations with VIMOS/VLT and data reduction. Section \ref{section:methods} describes the methods to derive galaxy properties used throughout the paper. In Section \ref{section:results} and Section \ref{section:discussion} we show and discuss the results from both individual and stacked spectral properties. Finally, Section \ref{section:conclusions} presents the conclusions. We use AB magnitudes, a Chabrier \citep{chabrier2003} initial mass function (IMF), and assume a $\Lambda$CDM cosmology with H$_{0}$=70 km s$^{-1}$Mpc$^{-1}$, $\Omega_{M}$=0.3, and $\Omega_{\Lambda}$=0.7. The physical scale at the redshift of the superstructure ($z\sim0.84$) is 7.63 kpc/\arcsec.


\begin{table*}
\caption{Observing log for our observations with VIMOS on the VLT for programmes 086.A-0895, 088.A-0550, and 090.A-0401 (PI: Sobral). The last two columns show the number of targeted objects for each pointing with a spectroscopic redshift and the spectroscopic success rate, respectively.}
\label{tab:observations}
\begin{tabular}{cccccccccc}
\hline
Pointing 	&	 R.A.	&	Dec. & Exp. time	& Dates  & Seeing & Sky & Moon & $N_\mathrm{z_\mathrm{spec}}$ & \% with $z_\mathrm{spec}$ \\
   &  (J2000)   	&	(J2000) 	& (ks)	 & (2013)  & ($''$)	& & & & \\
\hline
COSMOS-SS1 & 10 01 49  & +2 10 00  &	14.4  & Apr 14-16  &	 0.9  & Clear & Dark & {133} & {73\%} \\
COSMOS-SS2 & 10 01 33 & +2 10 00 &	14.4 & Apr 4-5, 8   &	0.8  & Clear & Dark  & {116} & {70\%}	\\
COSMOS-SS3 & 10 01 49 & +2 05 30 &     14.4 &  Apr 18; May 3-4	 &	0.9 & Clear & Dark & {110} & {74\%}	\\
COSMOS-SS4 & 10 01 33  & +2 05 30 &	14.4  &  Apr 5, 9, 12	&	0.8 &  Clear & Dark & {115} & {71\%} \\
COSMOS-SS5 & 10 01 49  & +2 00 00 &	14.4  &  Apr 15-17 	&	0.9 &  Clear & Dark & {117} & {71\%} \\
COSMOS-SS6 & 10 01 33  & +2 00 00 &	14.4  &  May 5, 7, 8, 11	&	0.9 &  Clear & Dark & {105} & {67\%}  \\
\hline
\end{tabular}
\end{table*}

\section{Sample and observations}\label{section:data}

\begin{table*}
\centering
\caption{Properties of the clusters in and around the VIMOS target fields (see Figure \ref{fig:cluster_region}). The cluster coordinates are from the catalogue produced by \citet{finoguenov2007}. The other properties were computed by \citet{balogh2014}. The third column is the median redshift of galaxy members. The fourth column is the intrinsic velocity dispersion. The fifth and sixth columns are the \textit{rms} projected distance of all group members from the centre and corresponding mass of the cluster, respectively.}
\label{tab:clusters}
\begin{tabular}{ccccccc}
\hline
{Label} & R.A.	&	Dec. & z	& $\sigma_i$   & $R_\mathrm{rms}$ & $M_\mathrm{rms}$\\
 & (J2000)   	&	(J2000) & 	& (km/s)	  & (Mpc) & ($10^{13}M_\odot$) \\
\hline
{A}&150.505 & 2.224 & 0.84 & $560\pm60$  & $0.81\pm0.07$ & $17.4\pm5.9$\\
{B}&150.370 & 1.999 & 0.83 & $420\pm40$ & $0.34\pm0.03$ & $4.2\pm1.3$\\
{C}&150.211 & 2.281 & 0.88 & $680\pm70$ & $0.23 \pm 0.03$& $7.5\pm2.8$\\
\hline
\end{tabular}
\end{table*}

	\subsection{The COSMOS superstructure at $z\approx0.84$}

By conducting a relatively wide ($\sim0.8$ square degrees) and deep (down to a flux limit of $8\times10^{-17}\mathrm{erg\ s^{-1}cm^{-2}}$) H$\alpha$ survey at $z=0.84$ in the COSMOS field, \cite{sobral2011} found a strikingly large overdensity of H$\alpha$ emitters within a region that happens to contain 3 X-ray clusters (first reported in \citealt{finoguenov2007}), as shown in Figure \ref{fig:cluster_region}. Limited spectroscopic observations from zCOSMOS \citep[][]{lilly2007} allowed to securely place the most massive cluster in the region at $z=0.835$, but the full structure seemed to span $z\approx0.82-0.85$ north to south. The H$\alpha$ imaging reveals a strong filamentary structure which seems to be connecting at least 3 cluster regions, but there are other possible groups/smaller clusters within the region \citep[][]{sobral2011}. Such structures around a massive cluster are similar to those found in other superstructures at $z\sim0.5-0.8$ \citep[e.g.][]{sobral2011,darvish2014,darvish2015,iovino2016}. Given the opportunity to study such a range of environments in a single data-set, we have designed a spectroscopic survey over this full region.

\begin{figure*}
\centering
\includegraphics[width=\linewidth]{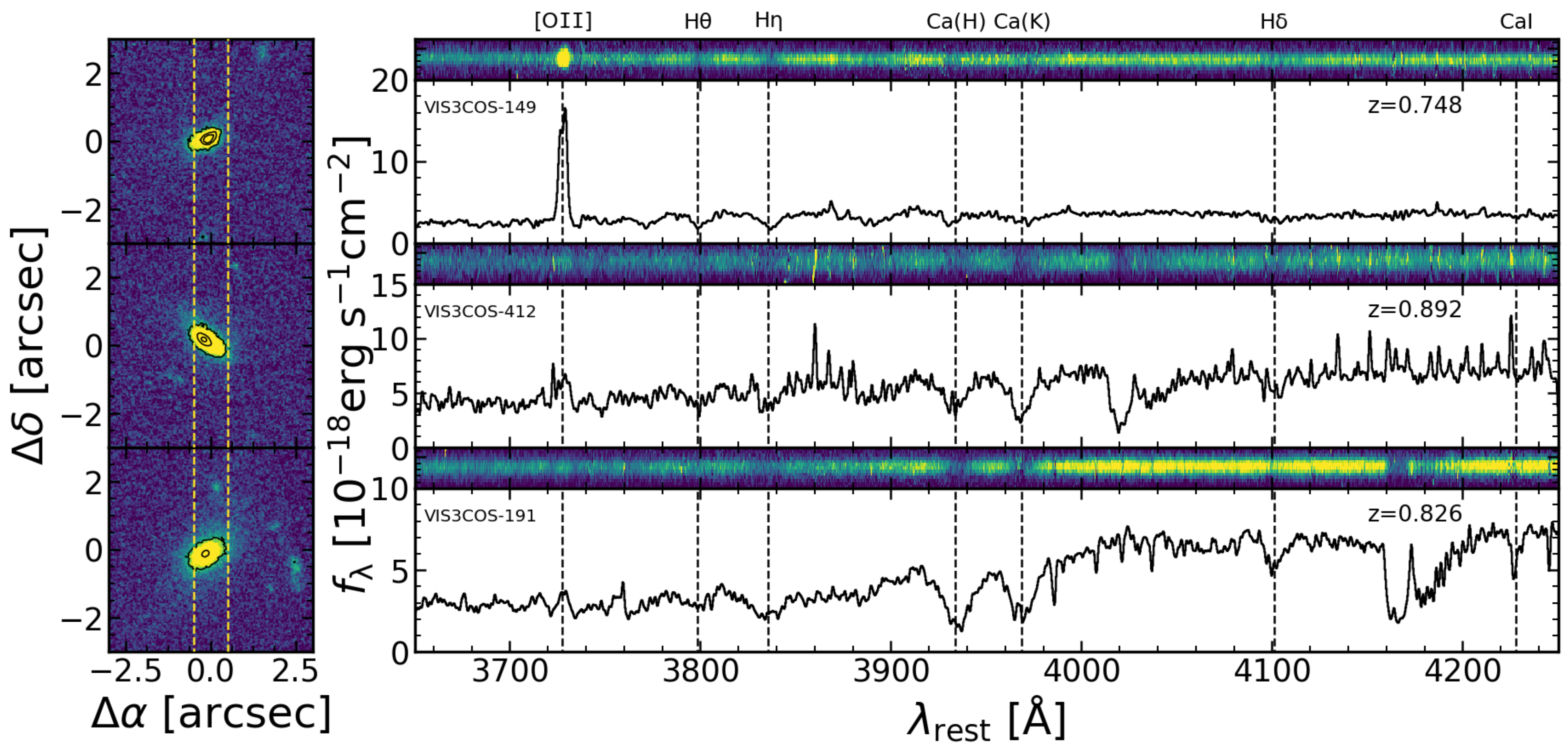}
\caption{Three individual examples of images and spectra obtained with our survey. Each thumbnail (left panel) shows the HST/ACS F814W image of each target from the COSMOS survey \citep{koekemoer2007} with the VIMOS slit overlaid (dashed yellow line). To the right of each stamp we show the corresponding 2D (top) and 1D (bottom) spectrum. We mark with vertical dashed lines the position of some spectral features present in our spectra.}
\label{fig:spectra_examples}
\end{figure*}

	\subsection{Target selection}\label{ssection:selection}

In order to accurately map the 3D large-scale structure at $z=0.84$ and identify the bulk of cluster, group, filament, and field members, we have targeted member candidates (using the VIMOS Mask Preparation Software to maximize the number of targets per mask) down to $I = 22.5$ (corresponding to stellar masses of $\approx$10$^{10}$\,M$_{\odot}$ for older and passive galaxies but much lower for younger galaxies, which have lower M/L ratios - see e.g. \citealt{sobral2011}). Our targets are selected by using state-of-the-art photometric redshifts (photo-$z$s) in COSMOS, using up to 30 narrow, medium, and broad bands \citep[c.f.][]{ilbert2009}. In practice, we use the upper and lower limits of the 99\% photo-$z$ confidence interval and select all sources for which such interval overlaps with $0.8<z<0.9$ (including sources best-fit by a quasar/AGN template). We reject all sources that are likely to be stars by excluding those sources for which $\chi^2(\rm star)/\chi^2(galaxy)<0.2$ \citep[c.f.][]{ilbert2009} or with clear star-like morphologies in high-resolution HST imaging and presenting near-IR vs. optical colours, which clearly classifies them as stars \citep[following e.g.][]{sobral2013}.

In order to effectively fill the masks, we introduce galaxies down to $I = 23.0$ and with photo-$z$s of $0.6<z<1.1$. We note that we use the 99\% photo-$z$ confidence interval instead of the best photo-$z$ to avoid significant bias towards redder and older galaxies (as blue and younger galaxies tend to present the largest scatter in the photometric versus spectroscopic redshift comparison). We also note that this selection recovers all our blue and star-forming H$\alpha$ emitters \citep[][]{sobral2009,sobral2011}. We can therefore fully map the supercluster without major selection biases. \add{In total, out of our entire parent sample of 1015 primary targets and 2257 secondary targets, we have placed 531 ($\sim$55\% of the parent primary) slits on primary targets and 440 ($\sim$19\% of the parent secondary) on secondary targets. Due to the six pointings targeting the same area we are not substantially biased against targets in higher densities (see also Section \ref{sec:completeness}).} Observations are described in Section \ref{ssection:observations}. We discuss our sample completeness in terms of spectroscopic success and relative to our parent sample in Appendix \ref{sec:completeness} and apply corrections whenever completeness effects might bias our results (see example Section \ref{ssection:sfr}).

	\subsection{Observations}\label{ssection:observations}

We have targeted the COSMOS superstructure identified in \cite{sobral2011} and studied photometrically in e.g. \cite{darvish2014}. We have used the High Resolution Red grism (HR-Red) with VIMOS {\citep{lefevre2003}} and the GG475 filter\footnote{This is the same mode used by LEGA-C, see \cite{vanderwel2016} for more details.}. Our observations are summarized in Table \ref{tab:clusters} and probe the rest-frame $3400-4600$\,\AA\ for our main targets (at $z\sim0.8$) with an observed 0.6\,\AA\ pix$^{-1}$ spatial scale, which at $z\sim0.8$ is $\sim0.33$\,\AA\ pix$^{-1}$ rest-frame. This allows for a clear separation of the spectral features and very accurate redshift determinations. Spectra cover a key spectral range at $z\approx0.84$, from [O{\sc ii}]\,$\lambda$3726,$\lambda$3729 (partially resolving the doublet, as our resolution is $\sim1$\,\AA\ for $z\sim0.8$ sources) through 4000\,\AA\ (allowing us to measure D4000, see Figure \ref{fig:spectra_examples}) to beyond H$\delta$ at high resolution (allowing to measure many other absorption lines and obtain their width).

The observations cover a contiguous overdense region of $21\arcmin\times31\arcmin$ ($9.6\times14.1$ Mpc, see Figure \ref{fig:cluster_region}) using 6 VIMOS pointings (chosen to overlap in order to assure both a contiguous coverage and a good target coverage and completeness, particularly for sources located in the densest regions). We have used the VIMOS 1{\arcsec} width slit with an average of 9{\arcsec} slit length. Our setup allowed us to offset different observing blocks by $\pm1.3${\arcsec} along the slit to guarantee an optimal sky subtraction. Observations were conducted in service mode in April and May 2013 (see Table \ref{tab:observations}) under clear conditions, new moon and an average seeing of 0.9{\arcsec} (ranging from 0.6{\arcsec} to 0.95{\arcsec}). Our pointings, labelled COSMOS-SS1 through COSMOS-SS6, have a total exposure of 4 hours each. Arcs and flats were taken each night. See Table \ref{tab:observations} for further details.

	\subsection{Data reduction}\label{ssection:dataredux}

Data reduction was done using the VIMOS ESO pipeline, version 6.10, through gasgano\footnotemark{}\footnotetext{http://www.eso.org/sci/software/gasgano.html}. The reduction is performed quadrant by quadrant (VIMOS has 4 different quadrants, labelled Q1 through to Q4). First, a master bias per night of observations is created by median combining bias frames per quadrant. Appropriate recipes are run in order to create master flats and master arcs for wavelength calibration. The pipeline is used to flag and mask hot pixels and cosmic rays and also to distort correct the observations. We obtain a sky subtracted spectra by estimating the median sky emission in several apertures away from each extracted source. Finally, 2D spectra are obtained by combining spectra obtained over different observing blocks. The extraction of the 1D spectra is conducted by collapsing the spectra in wavelength and then extracting along the trace's FWHM. We obtain our 2D and 1D spectra with a relative flux calibration. We are able to extract 1D spectra for 971 sources, with varying levels of S/N. See Figure \ref{fig:spectra_examples} for examples of individual 1D and 2D spectra.

	\subsection{Flux calibration}\label{ssection:fluxcalib}

Due to the wealth of available well-calibrated photometry for all our sources, we use broad and medium band data from COSMOS to test and then scale the flux calibration of our spectra. This allows us to obtain more accurate flux calibrations and to slit correct more appropriately than using a single standard star for each quadrant. This also allows us to correct for any misalignment in the slit position relative to each source. 

Briefly, we use the $I$-band selected photometric catalogue presented by \cite{ilbert2009} and start by using the $I$-band magnitudes. We convert $I$-band magnitudes into flux densities for each of our targets and compare those with the integral of the spectra convolved with the $I$-band filter. We then scale each spectra by the appropriate flux normalisation such that the integral within the $I$-band filter equals the flux density derived from photometry. Note that it also allows us to obtain a relatively good slit correction and thus we do not apply any further slit corrections for our data. For galaxies which are too faint in the $I$-band, we use the median flux calibration for the pointing and quadrant it was observed in. \add{This flux calibration is done under the assumption that galaxies have an homogeneous colour over their extent.}

As a further check, we also use the COSMOS medium band flux densities \citep[see e.g.][]{ilbert2009} and check that our flux calibration is valid for the full range of available medium bands. We find very good agreement at all wavelengths within $\pm10-15$\% which we interpret as our uncertainty in the flux calibration.

\subsection{Redshift measurements}\label{ssection:redshifts}

\begin{figure}
\centering
\includegraphics[width=\linewidth]{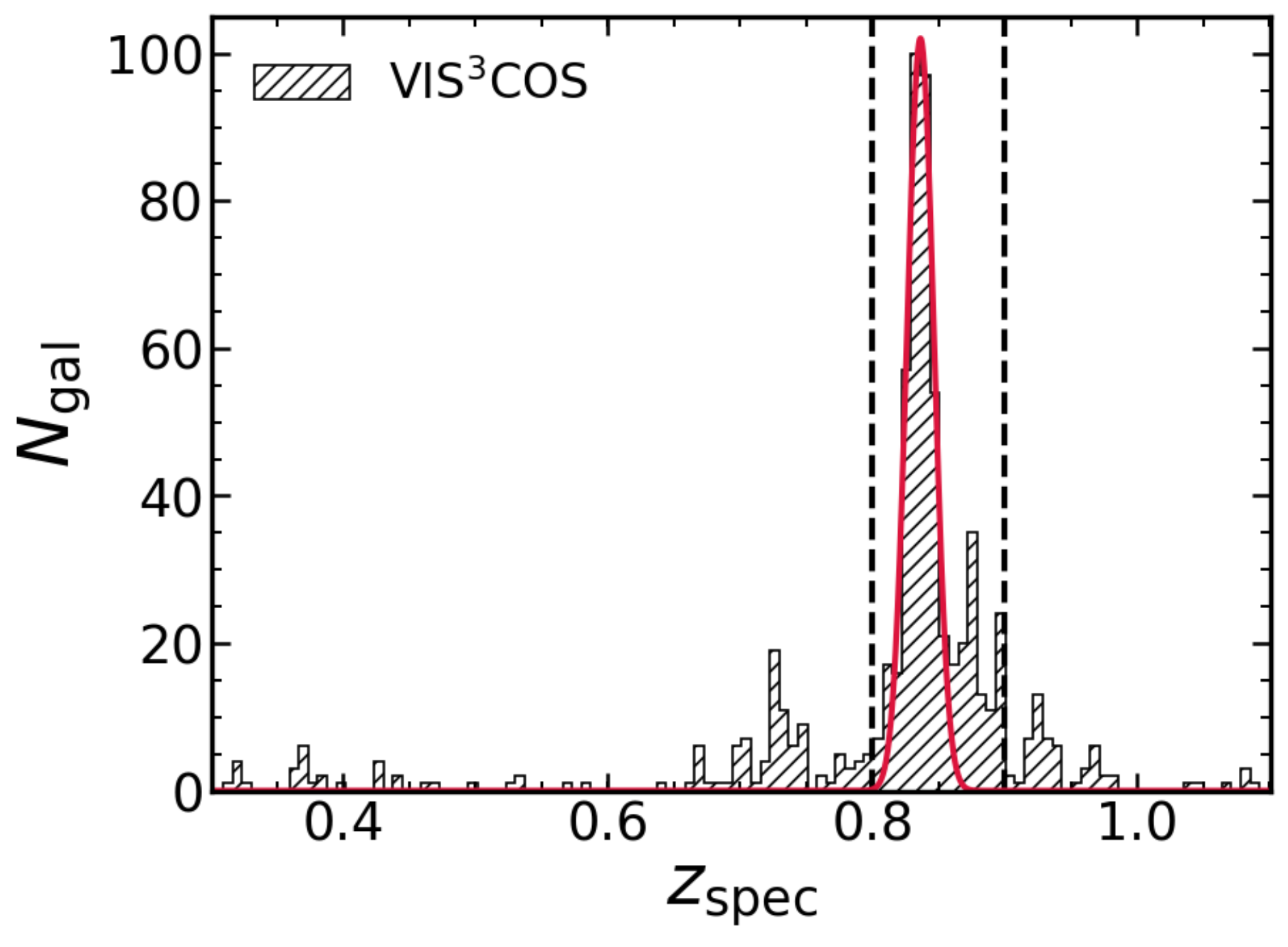}
\caption{Spectroscopic redshift distribution of the galaxies targeted in our sample as a black histogram. The vertical black dashed lines delimit the redshift selection of the results presented in this paper. The red line shows our Gaussian fit to the distribution \add{without using rejection algorithms}, pinpointing $z=0.836\pm0.008$ as the core redshift of the densest structure we find. The peak at slightly higher redshift ($z\sim0.88$) is likely produced members from the north-western cluster C (see Table \ref{tab:clusters} and Figure \ref{fig:cluster_region}).}
\label{fig:zspec_dist}
\end{figure}

We use the 1D spectra to measure accurate redshifts using \textsc{SpecPro} \citep{masters2011} and identify the bulk of the superstructure members. Most redshifts are derived from a combination of H+K absorption and other dominant absorption features such as the G-band for passive galaxies, while for star-forming galaxies we can detect [O{\sc ii}]\,$\lambda$3726,$\lambda$3729, in addition to absorption features. For a fraction of galaxies, we detect other lines such as H$\delta$ (in either absorption or emission). Redshifts are obtained by visually inspecting all spectra one by one and by searching the features mentioned above. We obtain secure redshifts for 696 sources with high S/N. The redshift distribution for the galaxies in our sample is shown in Figure \ref{fig:zspec_dist}.

\subsection{Final sample}

Our final sample is restricted to be at  $0.8<z<0.9$ to match our primary selection (see Section \ref{ssection:selection}) and has a total of 490 galaxies spanning a large diversity of environments across several Mpc that contain 3 X-ray confirmed galaxy clusters. We are releasing the final catalogue with this paper and we show in Table \ref{table:Gal_props} the first 10 entries.


\section{Determination of galaxy properties}\label{section:methods}

	\subsection{Measurement of [O{\sc ii}]\,$\lambda$3726,$\lambda$3729 line}\label{ssection:linemeasures}

\add{To obtain flux measurements of emission lines from our spectra, we interactively iterate through the entire dataset and zoom to a window of ${100}$\,\AA\ around {[O{\sc ii}]\,${\lambda}$3726,${\lambda}$3729}. We define two regions of ${\sim15}$\,\AA\ (one blueward, one redward of the line) from which we estimate the median continuum level. Then the local continuum is defined as a straight line that goes through those points. To fit the doublet we use a combination of two Gaussian models through the functional form:
 
\begin{equation}
{f(\lambda) = A_1 \exp\left[- \frac{(\lambda - \lambda_1)^2}{2\sigma^2} \right] +  A_2\exp\left[- \frac{(\lambda - \lambda_2 )^2}{2\sigma^2} \right],}
\label{eq:guassian_fit}
\end{equation}

\noindent
with 3 free parameters: ${A_1,A_2}$, and ${\sigma}$. The parameters ${A_1}$ and ${A_2}$ are the amplitudes of each component, ${\sigma}$ is the width of each Gaussian component. The centre of each component is fixed to be ${\lambda_1=3726.08\pm0.3}$\AA\ and ${\lambda_2=3728.88\pm0.3}$\AA\ (we allow for a small shift in the line centre that is of the size of the resolution element of the spectra). To estimate the line properties we use the information on the error spectra and perturb each flux at all wavelengths considered for the fitting by drawing a random number  on the observed value and with a width that is equal to its error. We run this exercise 10000 times and then estimate the errors on the line fit by taking the 16th and 84th percentile of the distribution in each free parameter. 

From now on we only use individual measurements if the S/N is ${>3}$. We note that in Section \ref{ssection:specstacks} we will obtain and measure stacks as a function of environment, allowing us to obtain the median properties of spectral lines for specific subsets of galaxies irrespective of their individual detection. This of course leads to a much higher S/N. We measure the line properties of the stacks with the same procedure described here for individual sources.}

\begin{figure}
\centering
\includegraphics[width=\linewidth]{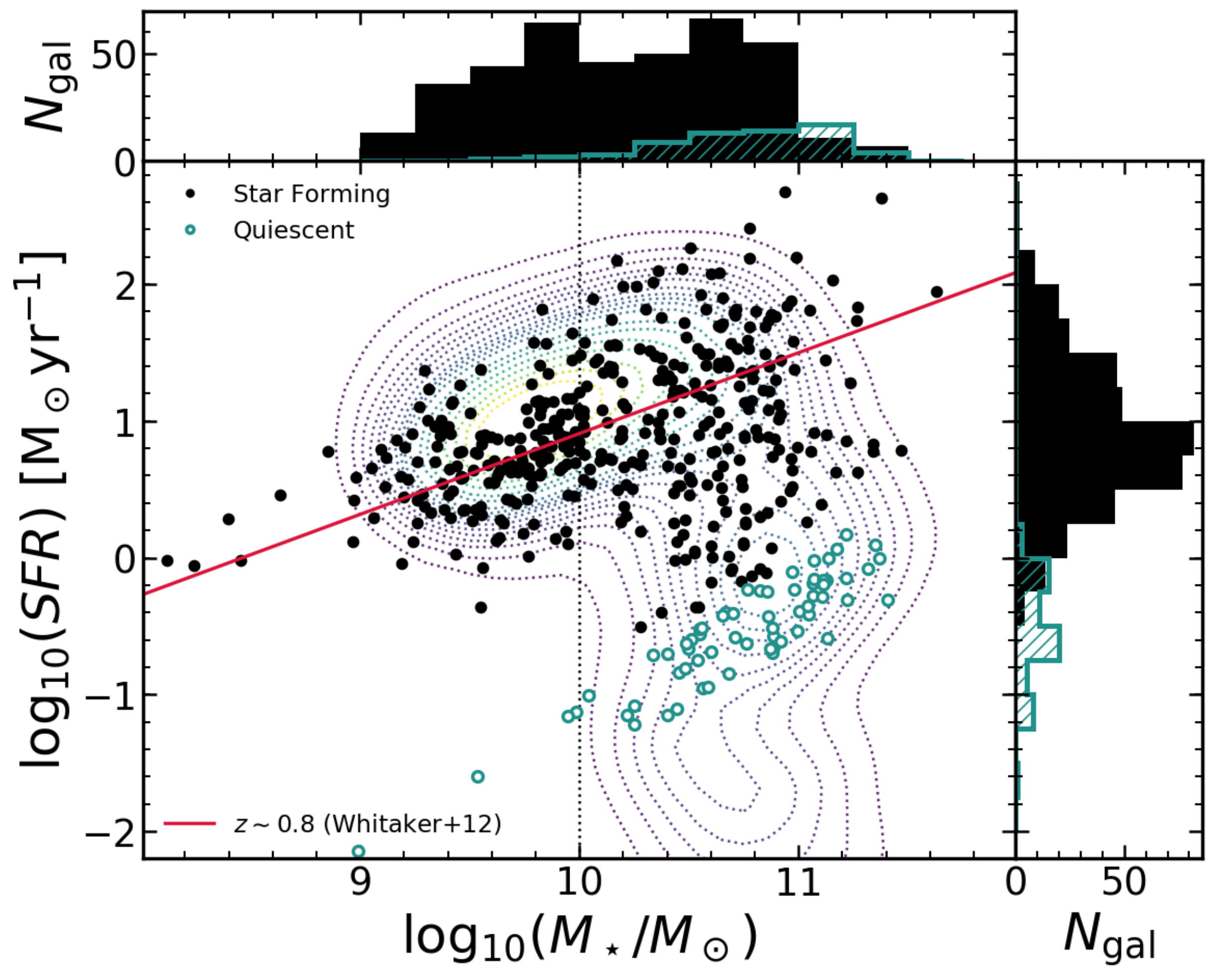}
\caption{Stellar masses and star formation rates derived from SED fitting (see Section \ref{ssection:sedfits}) in our spectroscopic sample at $0.8<z<0.9$. For comparison we show the derived best-fit relation for star-forming galaxies computed at $z=0.84$ using the equation derived by \citet{whitaker2012} over a large average volume in the COSMOS field. The vertical dotted line shows the completeness limit of our survey. The dotted contours show the COSMOS2015 \citep{laigle2016} distribution of galaxies with $0.8<z_\mathrm{phot}<0.9$ and $i_\mathrm{AB}<23$ from 10\% to 95\% of the sample in 5\% steps. Empty circles highlight the photometric quiescent sample with $\log_{10}(sSFR)<-11$.}
\label{fig:mass_sfr_dist}
\end{figure}

	\subsection{Stellar masses and star formation rates}\label{ssection:sedfits}

To estimate the stellar masses and star formation rates for the galaxies in our sample, we have performed our own Spectral Energy Fitting (SED) using \textsc{magphys} \citep{cunha2008} and our knowledge of the spectroscopic redshift to better constrain the range of possible models. The models were constructed from the stellar libraries by \citet{bruzual2003} using photometric bands from near-UV to near-IR (Galex NUV, Subaru uBVriz, UltraVISTA YJHK$_s$, SPLASH-IRAC 3.6$\mu$m,4.5$\mu$m,5.8$\mu$m,8$\mu$m) taken from the COSMOS2015 photometric catalogue \citep{laigle2016} and the dust absorption model by \citet{charlot2000}. We found COSMOS2015 matches for 466 out of the 490 galaxies that are in our selected redshift range between $0.8<z<0.9$ for which we obtained the physical parameters that we use throughout the paper (stellar mass and star formation rates). If not found in the COSMOS2015 catalogue we do not obtain any estimate for stellar mass and SFRs through SED-fitting, which happens only for 3\% of the sample. These missing sources are serendipitous objects which are faint in the $I$-band and below our completeness limit. We compare our results on stellar mass and star formation rates with those provided in the COSMOS2015 catalogue and find a dispersion of $\sim0.3$\,dex for the stellar mass and $\sim0.7$\,dex for the SFRs.

We present in Figure \ref{fig:mass_sfr_dist} the stellar masses and star formation rates in our sample and show that we are probing galaxies with $\log_{10}\left(M_\star/\mathrm{M_\odot}\right)\gtrsim9$ in a wide range of SFRs ($-2\lesssim\log_{10}\left(SFR\right)\lesssim2$). We see that our sample includes normal star-forming galaxies  as well as galaxies that are found well below the SFR main sequence \citep[see e.g.][]{noeske2007,elbaz2007,whitaker2012}, which are characteristic of galaxies in the process of star formation quenching or just quenched \citep[e.g.][]{fumagalli2014}. To select quiescent galaxies within our sample we impose a specific SFR cut at $\log_{10}(sSFR)<-11$ \citep[see e.g.][]{ilbert2010,carollo2013} and find a total of 64 galaxies in these conditions.

We also obtain from \textsc{magphys} the effective optical depth of the dust in the $V$-band, $\tau_V$, which we translate into an average reddening value of $E(B-V) = 1.086\tau_V/R_V$ \citep[assuming $R_V=3.1$, see e.g.][]{draine2004}. We find that our galaxies have an average reddening value of $E(B-V)\sim0.27\pm0.02$. We report here that above $\log_{10}\left(M_\star/\mathrm{M_\odot}\right)\gtrsim10$ there is little dependence of the median extinction with stellar mass with a reddening value of $E(B-V)\sim0.32\pm0.02$ ($\sim0.37\pm0.02$ if we consider star-forming only).

\add{We measure the [O{\sc ii}] line flux by integrating over the best fit model described by Eq. \ref{eq:guassian_fit}, which can be solved analytically as ${F = \sigma\sqrt{2\pi}\left(A_1+A_2\right)}$}. We correct the measured [O{\sc ii}] luminosity by the SED extinction value. The corrected luminosity is given by

\begin{equation}
{L_\mathrm{[OII],corr} = L_\mathrm{[OII]}/e^{-\tau_\mathrm{[O\mathtt{II}]}},}
\end{equation}

\begin{figure}
\centering
\includegraphics[width=\linewidth]{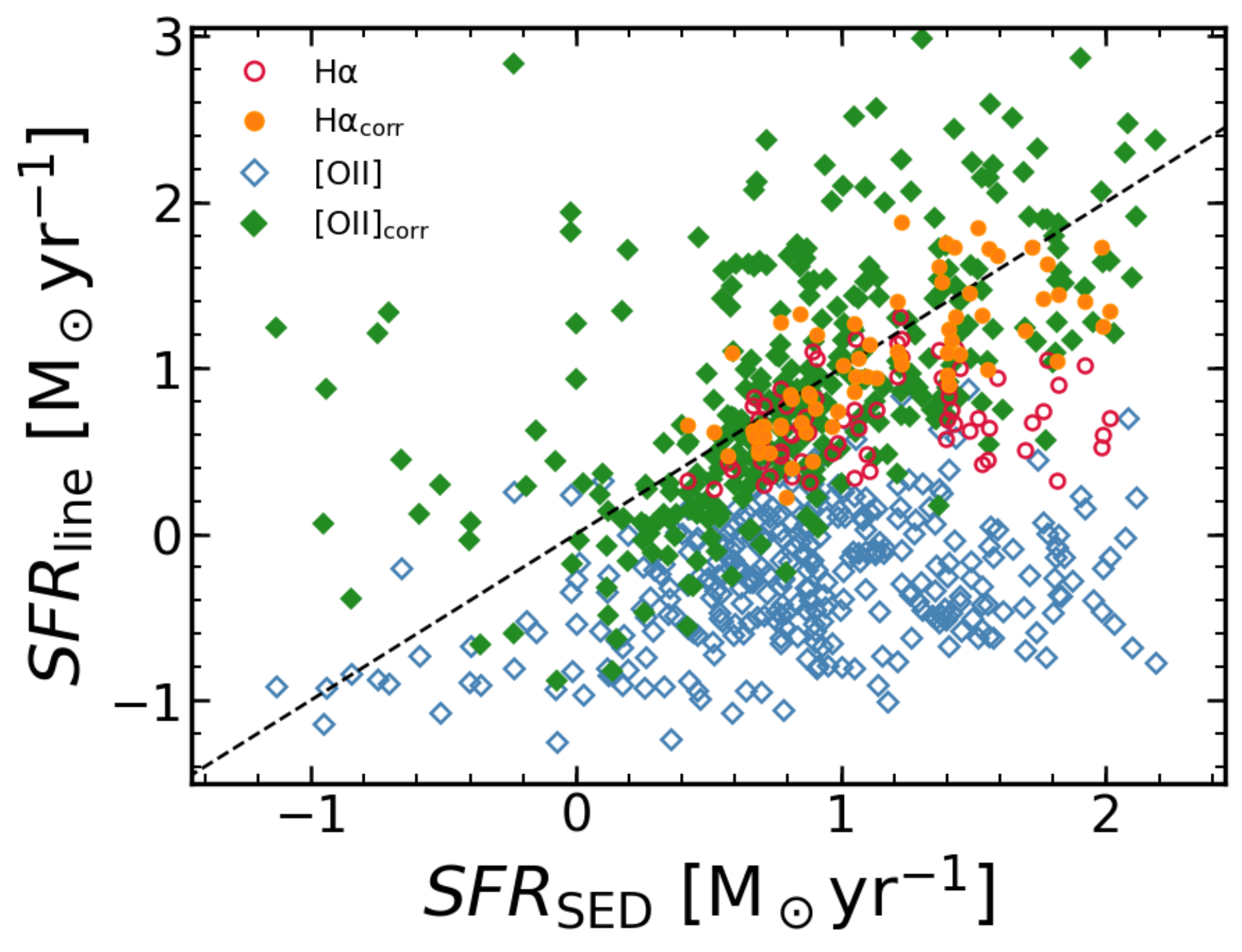}
\caption{SFR estimates from SED fitting and from [O{\sc ii}] (derived from equation 4 of \citealt{kewley2004}) of the galaxies in our spectroscopic sample at $0.8<z<0.9$. As red circles we show the dust uncorrected H$\alpha$ derived star formation rates for the galaxies in our sample and that were measured by \citet{sobral2011}. The subscript \emph{corr} denotes the dust corrected SFRs derived from each estimate using the optical depth derived through SED fitting (see Section \ref{ssection:sedfits}).}
\label{fig:sfr_comp}
\end{figure}

\noindent
\add{where ${\tau_\mathrm{[O\mathtt{II}]}}$ is the optical depth at ${\lambda=3727}$\AA\ derived using the dust model used in \sc{magphys} \citep{charlot2000}.} The effect of extinction on the luminosity of [O{\sc ii}] is displayed on Figure \ref{fig:sfr_comp} and it can account for the difference that we find when comparing SED and [O{\sc ii}] SFRs using \add{the calibration derived by \citet{kewley2004} and applying a conversion factor between \citet{salpeter1955} and \citet{chabrier2003} IMFs}:

\begin{equation}
{SFR = \frac{6.58\times10^{-42}}{1.7}L_\mathrm{[OII],corr}.}
\label{eq:SFR_OII}
\end{equation}

We find a spread of \add{0.64\,dex but on average the derived SFRs are consistent with each other (median difference of 0.07\, dex)}. We also show the SFR as derived from H$\alpha$ luminosity \citep{kennicutt1998} from the HiZELS survey, which was used to first pinpoint the existence of this structure \citep{sobral2011}. 

\add{We stress however that [O{\sc ii}] emission can originate from other sources not related to star formation (e.g. AGN, LINERs) and that it is a poor tracer of SFR for red galaxies \citep[e.g.][]{yan2006,kocevski2011}. This tracer is also dependent on the metallicity of the galaxy \citep{kewley2004}. Those are the reasons for our choice to do our analysis in terms of star formation in galaxies using the quantity derived from SED fitting instead of relying on [O{\sc ii}] emission as a tracer of SFR.}

	\subsection{Overdensities estimation}\label{ssection:overdensities}

\begin{figure}
\centering
\includegraphics[width=\linewidth]{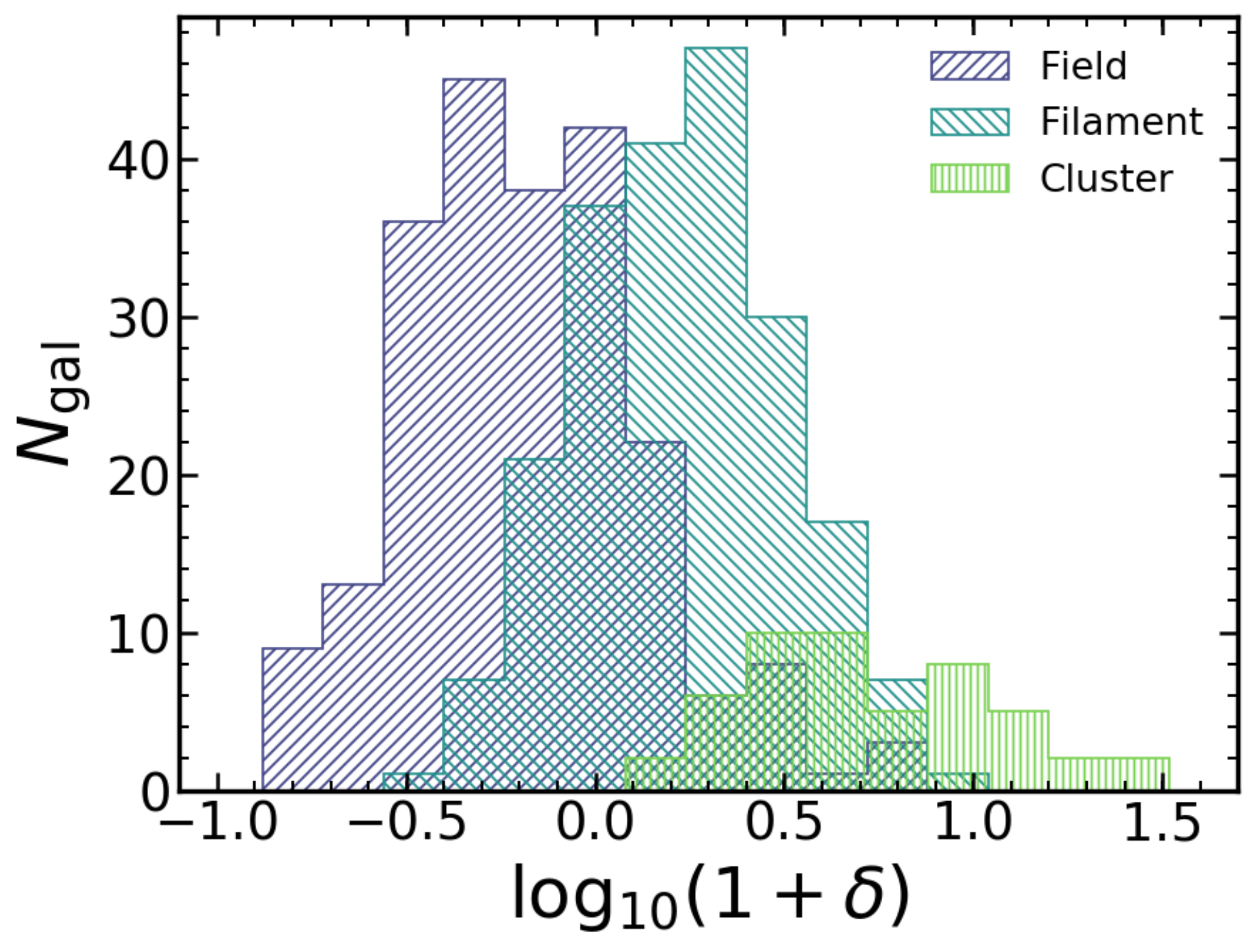}
\caption{Overdensity distribution for the galaxies in our sample with $0.8<z<0.9$. We show the different cosmic web environments of galaxies (field, filament, and cluster) according to their classification using the scheme devised by \citet{darvish2014,darvish2017}.}
\label{fig:dens_dist}
\end{figure}

The estimate of local overdensity was computed \add{as described by \citet{darvish2015a,darvish2017}} and is based on the photometric redshift catalogue of the COSMOS survey presented by \citet[][\!\!, {see also \citealt{muzzin2013,laigle2016}}]{ilbert2013}. \add{The density field was computed over an area of $\sim1.8\,\mathrm{deg^{2}}$ using a mass-complete sample with accurate photometric redshifts spanning ${0.1<z_\mathrm{phot}<1.2}$}. The surface density field was computed in 2D slices of redshift of \add{widths ${\pm 1.5\sigma_{\Delta z/(1+z)}}$ \citep[as suggested by][]{malavasi2016}}. To properly account for the uncertainty on the photometric redshift estimate, the full photo-$z$ PDF of each galaxy is taken into account. Then, at each redshift slice, we select all galaxies which fall in that slice and assigned it a weight corresponding to the percentage of the photo-$z$ PDF contained in that slice. \add{We use all galaxies which have weights greater than 10\% in the corresponding slice.} The surface density assigned to each point in the density field is (based on adaptive kernel smoothing):

\begin{equation}
\Sigma_i = \frac{1}{\sum_{i=1,}^{N}w_{i}} \sum_{i=1,}^{N} w_{i}K(\vec{r},\vec{r}_i,h_i),
\end{equation}

\noindent
where $\vec{r}$ is a location in the density field, $\vec{r}_i$ is the position of each galaxy, $w_{i}$ is the weight assigned to each galaxy, $h_{i}$ is the kernel width at the position of each galaxy, and $K$ is a two-dimensional Gaussian kernel function.

In these equations, $N$ is the number of galaxies in the slice \add{with weights greater than 10\%}, $r_i$ is the position of the galaxy, $r_j$ is the position of all other galaxies in the slice, and $h_i$ is the adaptive smoothing parameter for our assumed kernel. \add{The value of ${h_i=h\sqrt{G/\Sigma_i}}$ Mpc, where ${\Sigma_i}$ is the initial density estimation at the position of galaxy ${i}$ using a fixed kernel with of 0.5 Mpc width, ${G}$ is the geometric mean of all ${\Sigma_i}$ at each redshift slice, and ${h}$ is chosen to have a value around the typical size of X-ray clusters \citep[0.5 Mpc, see e.g.][]{finoguenov2007}. We then evaluate the density field in a 2D grid with a spatial resolution of 50 kpc at each redshift.} We define overdensity as:

\begin{equation}
1+\delta = \frac{\Sigma}{\Sigma_\mathrm{median}}
\label{eq:density}
\end{equation}

\noindent
with $\Sigma$ being the projected local density and $\Sigma_\mathrm{median}$ being the median of the density field of the redshift slice the galaxy is in. \add{We choose to use number densities instead of mass density estimates \citep[e.g.][]{wolf2009} to avoid introducing any bias due to any underlying relation between stellar mass and density that may exist.} For a more detailed description of the method, we refer the reader to \citet{darvish2014} and \citet{darvish2015a}.

We have computed the value of the overdensity \add{for each galaxy by interpolating the density field to their angular position and spectroscopic redshift}. We show in Figure \ref{fig:dens_dist} the distribution of our galaxies according to their overdensity and labelled by the region they are likely to belong to, as defined by the cosmic web measurements computed by \citet{darvish2014,darvish2017}. We note that when referring to galaxies within our spectroscopic sample in cluster regions we are mostly referring to either rich groups or the outskirts of massive clusters as our observational setup does not allow for a good sampling of densely populated regions due to slit collision problems.

We note that there is an overlap between the different labelled regions and the measured local overdensity in Figure \ref{fig:dens_dist}. This happens because the region assigned to each galaxy is based on the definition of the strength of the cluster and filament signals, which takes into account the morphology of the density field. That is the reason why a pure density-based definition of the environment of galaxies cannot fully separate them into real physical structures \citep[see e.g.][]{aragon-calvo2010,darvish2014}. This means for example that we can have dense filaments (as high density regions with thread-like morphology, likely infall regions of massive clusters) and less dense cluster regions (intermediate density with circular morphology, likely associated with galaxy groups). We refer to \citet[][see also \citealt{aragon-calvo2010}]{darvish2014,darvish2017} for more details.

	\subsection{Spectral stacks}\label{ssection:specstacks}

To increase the S/N on the obtained spectra and investigate details on the spectral properties of galaxies as a function of their stellar mass and local density, we have performed stacking of individual galaxy spectra. Our stacking method can be summarized as a median, interpolated, and normalized spectra. For each set of spectra we start by shifting the spectrum to its rest-frame wavelengths using the redshift we have measured (see Section \ref{ssection:redshifts}). Then we  linearly interpolate the spectra onto a common universal grid (3250-4500\,\AA, $\Delta\lambda=0.3$\,\AA/pixel). We normalize each spectrum to the mean flux measured from 4150-4350\,\AA. Lastly, we median combine all spectra by taking the median flux at each wavelength. We estimate that our typical errors in the spectroscopic redshift measurements are on the order of $\sim0.0005$, which translates to an error of $\sim1$ \AA, comparable to our spectral resolution at $z\sim0.8$. Thus, our stacking should not smear the lines enough to affect the measurements on the [O{\sc ii}]\,$\lambda$3726,$\lambda$3729 doublet.


\section{Results}\label{section:results}

Throughout this section, our measure of environment is quantified by $\delta$ (see Equation \ref{eq:density}). For a broad comparison between different environments, we defined as lower density galaxies residing in $\log_{10}(1+\delta)<0.1$ and as higher density galaxies residing in regions with \bm{\log_{10}(1+\delta)>0.4}.

	\subsection{Redshift distribution}\label{ssection:zdist}
	
\begin{figure}
\centering
\includegraphics[width=\linewidth]{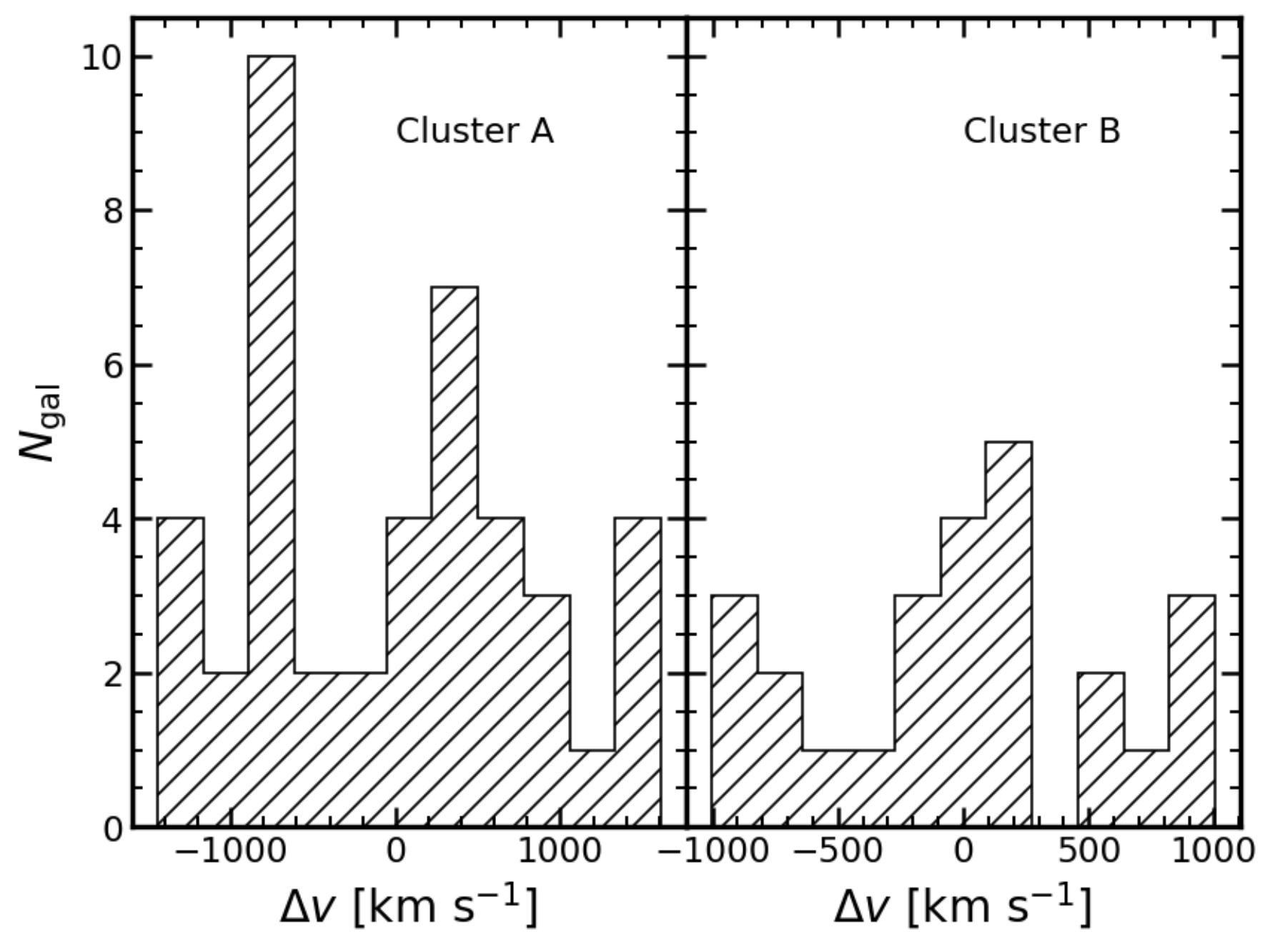}
\caption{\add{The velocity distribution for clusters A (left) and B (right) of all member galaxies. We note that these structures cannot be described by a single Gaussian shape indicating that these structures are not virialized.}}
\label{fig:clusterVelocities}
\end{figure}

From our first redshift measurements, based on 2-3 lines and the dispersion of the measurements, we are able to derive the full redshift distribution of our VIMOS sample. We show the results in Figure \ref{fig:zspec_dist} which shows a very clear peak at $z\approx0.84$. By fitting a Gaussian to the redshift distribution at $z\approx0.8$ we find that the COSMOS superstructure is well characterised by $z=0.84\pm0.01$ with 367 galaxies fully included within this redshift distribution. 

We attempt to estimate the mass of the two clusters for which we have coverage (A and B on Table \ref{tab:clusters}, see also Figure \ref{fig:cluster_region}) by computing the radial velocity dispersion, $\sigma_r$, of the spectroscopically confirmed galaxies in our sample. We estimate the size of the cluster by computing the root mean square of the distances, $R_\mathrm{rms}$, to the estimated centre (average position of selected members). We compute the velocity dispersion, $\sigma_r$, using the gapper technique \citep[][\!\!, see also \citealt{balogh2014}]{beers1990}. To get a final estimate for each cluster, we iterate 5 times and compute the mean position, $R_\mathrm{rms}$, and $\sigma_r$ by selecting at each step galaxies within 2$R_\mathrm{rms}$ of the cluster centre and within 2$\sigma$ of the median cluster redshift. We start our iteration procedure by assuming an initial guess for $R_\mathrm{rms}=0.5$ Mpc.

We find values of $\sigma_r=875\pm179$ $\mathrm{km\ s^{-1}}$ (43 galaxies) and $R_\mathrm{rms}=1.1$ Mpc for cluster A and of $\sigma_r=598\pm225$ $\mathrm{km\ s^{-1}}$ (25 galaxies) and $R_\mathrm{rms}=1.3$ Mpc for cluster B. Assuming a virial state for each cluster, we can estimate their mass as $M = 3\sigma_r^2R_\mathrm{rms}/G$. We find $M=6\pm3\times10^{14}\mathrm{M_\odot}$ and $M=3\pm2\times10^{14}\mathrm{M_\odot}$, respectively. These values are up to an order of magnitude higher than the values reported by \citet[][\!\!, see Table \ref{tab:clusters}]{balogh2014} and this difference is mainly driven by our larger derived values of $R_\mathrm{rms}$. We note here that our measurements are done under the assumption that the clusters are \add{virialized}. 
We hypothesize that when applying a similar criteria for galaxy membership as \citet{balogh2014}, we are likely picking up additional moving substructures (at slightly different redshifts) that are artificially increasing our measured cluster sizes and velocity dispersion. This is supported by the non-Gaussian shape of the velocity distribution histograms of the selected members \add{(see Fig. \ref{fig:clusterVelocities})}.

	\subsection{SFR dependence on local overdensity}\label{ssection:sfr}

\begin{figure}
\centering
\includegraphics[width=\linewidth]{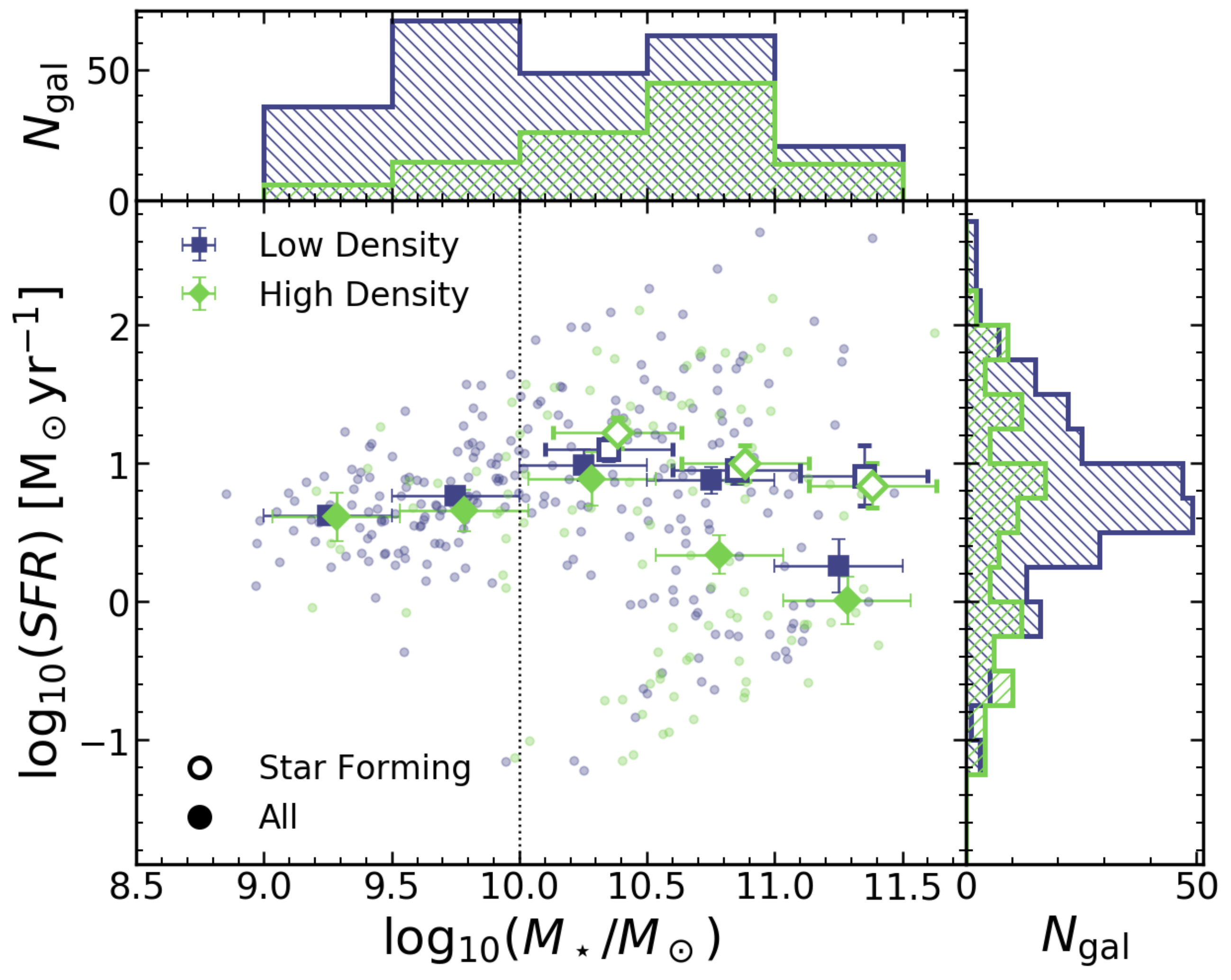}
\includegraphics[width=\linewidth]{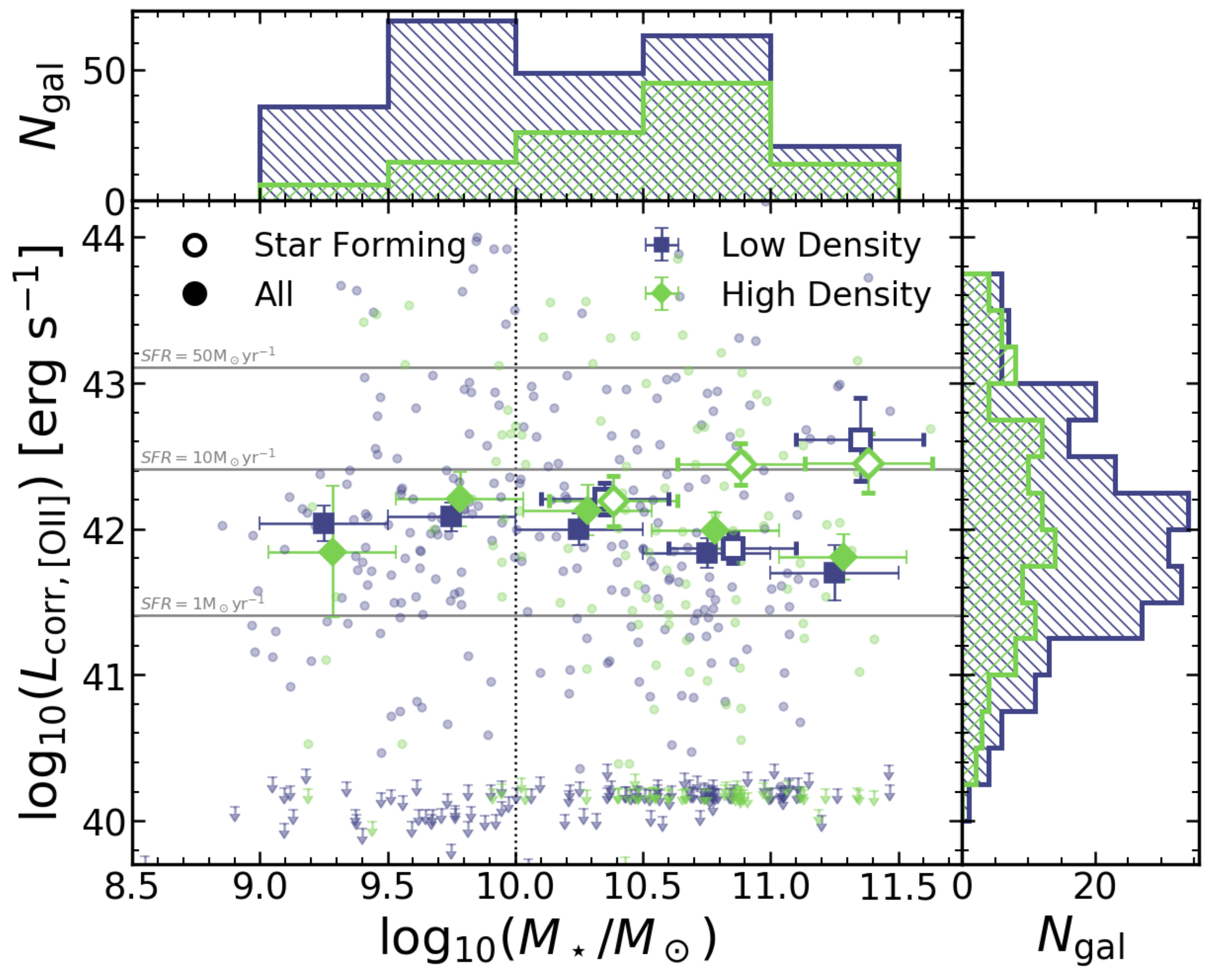}
\caption{{Top:} SFR (from SED fitting) distribution as a function of stellar mass. Each small circle represents a single galaxy. Large squares show the median value for the population in stellar mass bins. \add{Error bars show the error on the median of each bin}. Higher density regions are coloured in blue while low density galaxies are shown in green colours. The empty symbols represent the bins considering star-forming galaxies only, with $\log_{10}(sSFR)>-11$. The symbols are horizontally shifted for visualization purposes. The vertical dotted line shows the completeness limit of our survey. Globally, we find that galaxies in higher density regions have lower SFRs, but only when considering the entire population. When selecting star-forming galaxies, we find no difference between the median SFRs in low and high density environments. Bottom: Dust corrected [O{\sc ii}] luminosity distribution as a function of stellar mass. We show as small arrows the upper limits on [O{\sc ii}] luminosity for the galaxies which we have not a measure with sufficient S/N. \add{We show as horizontal lines three values of SFR = 1, 10, 50 ${\mathrm{M_\odot yr^{-1}}}$ as derived from Eq. \ref{eq:SFR_OII}.} We typically find no differences between low- and high-density regions in terms of the median dust-corrected [O{\sc ii}] luminosity at all stellar masses probed in our sample.}
\label{fig:SFR_Mass}
\end{figure}

\begin{figure}
\centering
\includegraphics[width=\linewidth]{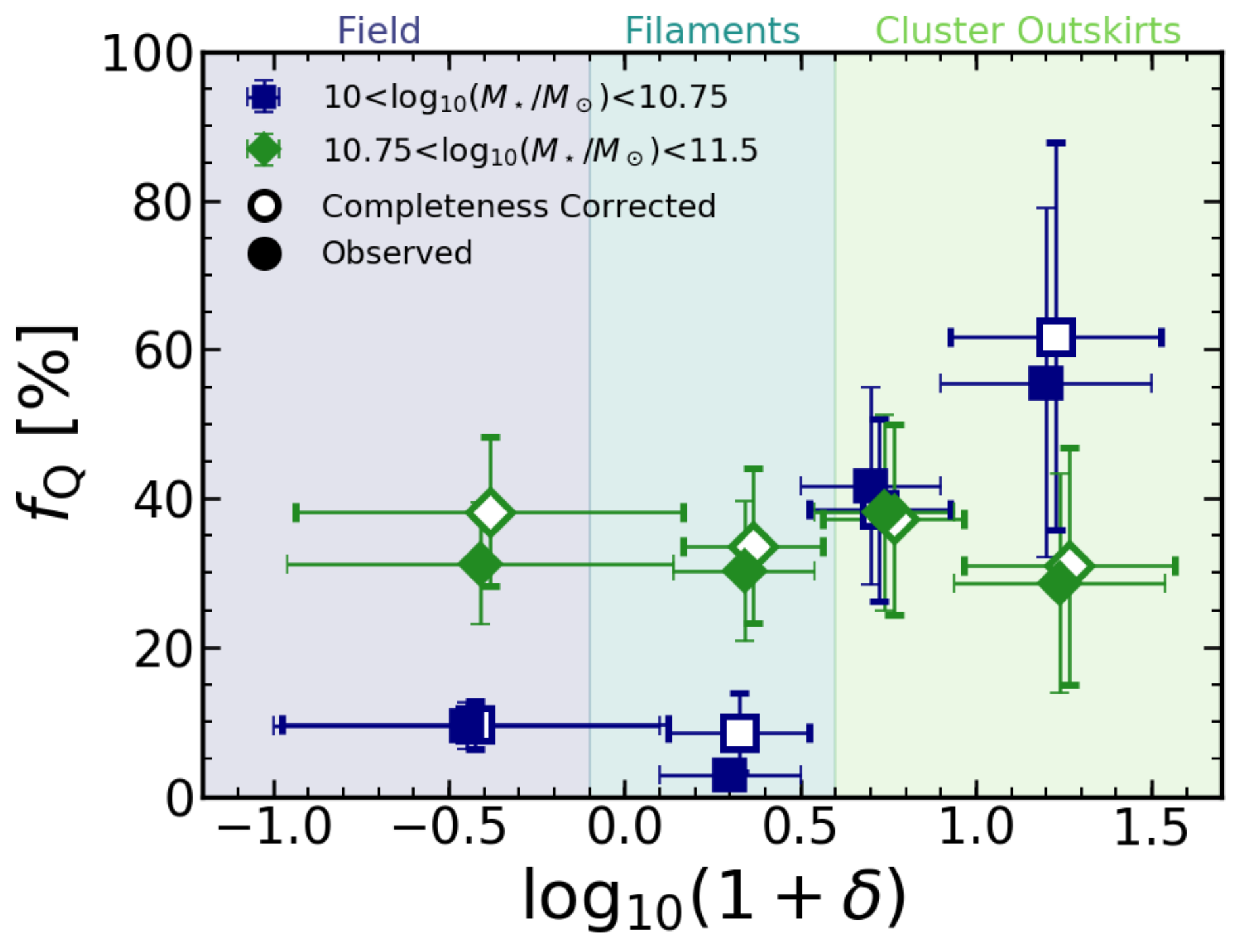}
\caption{The fraction of quenched galaxies within our sample ($f_\mathrm{Q}$, with $\log_{10}(sSFR)<-11$) as a function of local density and in two different bins of stellar mass. Open symbols show the same fraction after correcting for our sample completeness. \add{Error bars are computed using Poisson statistics.} We find in both cases that the lower stellar mass galaxies show a sharp increase for higher density environments whereas the highest stellar mass galaxies show no environmental dependence of $f_\mathrm{Q}$. \add{Shaded regions provide an approximate estimate of the cosmic web environment given the measured overdensity (but see Section \ref{ssection:overdensities} for more details).}}
\label{fig:SFR_Quench}
\end{figure}

We show in the upper panel of Figure \ref{fig:SFR_Mass} the dependence of SFR on stellar mass and local density. For low mass galaxies ($\log_{10}(M_\star/\mathrm{M_\odot})\lesssim10$) we find the same average SFR in both high and low density regions, although there are only very few ($\sim$10) low mass galaxies in our sample in high density regions and all of them are star-forming (check completeness in Figure \ref{fig:sample_completeness}). At higher stellar masses ($\log_{10}(M_\star/\mathrm{M_\odot})>10$), we find a stronger dependence of SFR on local density. At $\log_{10}(M_\star/\mathrm{M_\odot})\sim10.75$ the difference is the highest due to a larger fraction of galaxies at these stellar masses being photometrically defined as quiescent in higher density regions. At the highest stellar masses ($\log_{10}(M_\star/\mathrm{M_\odot})\gtrsim11$), there are few star-forming galaxies in both high and lower density regions and we see little dependence of the star formation activity in galaxies with the local density in which they reside. \add{We fit a linear model, ${\log_{10}(SFR) = m\log_{10}(M)+b}$, to the stellar mass-SFR relation at ${\log_{10}(M_\star/\mathrm{M_\odot})>10}$ and find that for lower density regions the null hypothesis of a flat relation (${m=0}$) is rejected at ${\sim3.3\sigma}$ (${m=-0.7\pm0.22}$) and for higher density regions is rejected at ${\sim3.5\sigma}$ (${m=-0.9\pm0.25}$).}

When looking at the trends considering star-forming galaxies only (with $\log_{10}(sSFR)>-11$), the difference between low- and high-density regions vanishes. \add{With a similar linear model as described above we find ${m=-0.2\pm0.23}$ for low-density regions and ${m=-0.4\pm0.20}$ for higher density regions. These models are less than 2${\sigma}$ from the null hypothesis.} When compared with the full sample, it suggests that the decline in the median SFR of the full sample in dense regions is mainly driven by the higher fraction of quenched galaxies.

We compute the fraction of galaxies that are defined as quenched in our sample ($\log_{10}(sSFR)<-11$) and show our findings in the Figure \ref{fig:SFR_Quench}. \add{Error bars for the fraction of quiescent galaxies are computed using Poisson statistics $\boldsymbol{\left(\Delta f_Q = f_Q\sqrt{N_{Q}^{-1}+N_{T}^{-1}-2N_{Q}^{-1}N_{T}^{-1/2}}\right)}$.} We inspect the environmental dependence of this fraction on environment for two separate stellar mass bins. We find that the lower stellar mass galaxies ($10<\log_{10}(M_\star/\mathrm{M_\odot})<10.75$) have a nearly constant quenched fraction at low to intermediate densities. We then find a jump from {$\sim10\%$ to $\sim40-60\%$} towards higher density regions. When considering the higher stellar mass bin ($\log_{10}(M_\star/\mathrm{M_\odot})>10.75$) we find no dependence of the quenched fraction on local density, being nearly constant at $\sim30\%$. We show also the reported values after correcting for our sample completeness and we find qualitatively the same results (see Appendix \ref{sec:completeness} for more details on the spectroscopic sample completeness).

\begin{figure*}
\centering
\includegraphics[width=\linewidth]{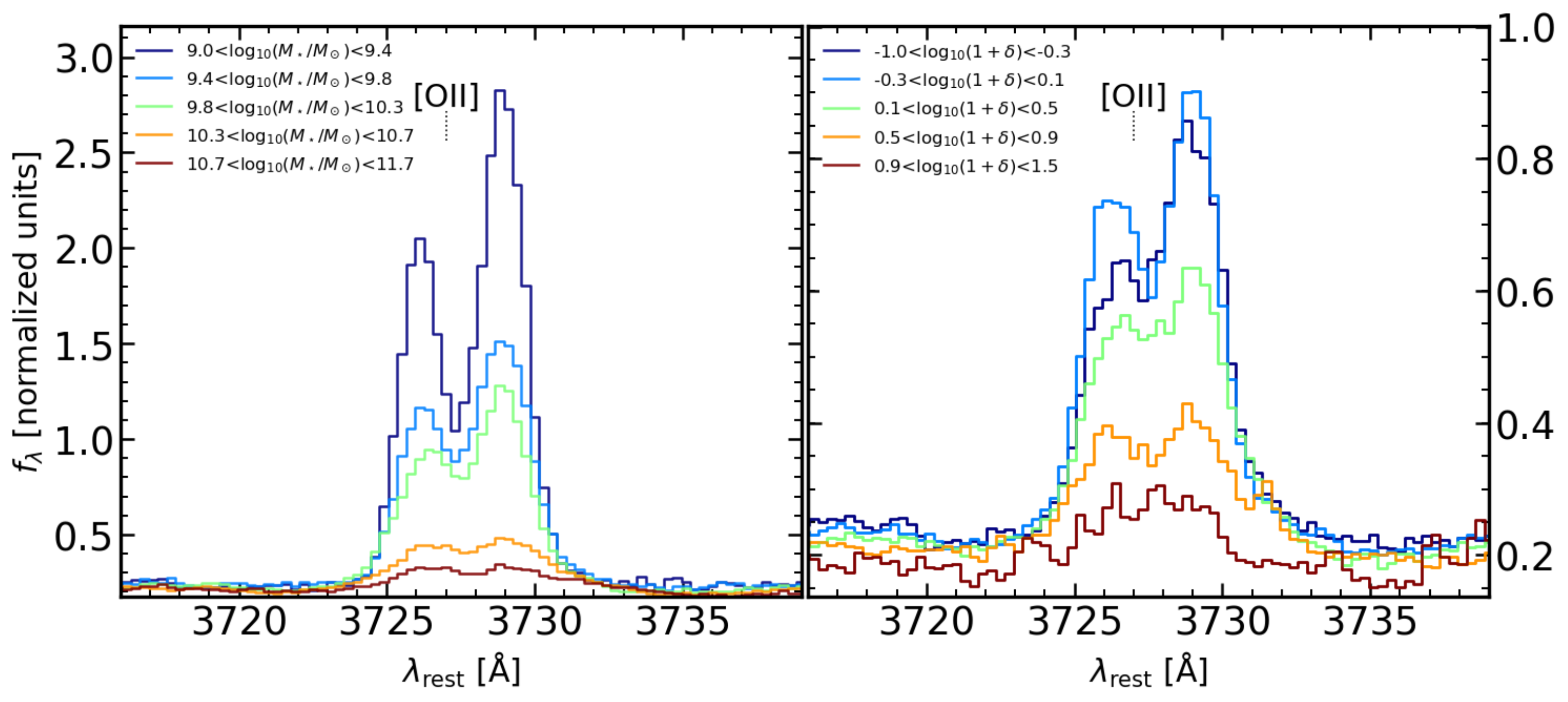}
\caption{Resulting spectral stacks (normalized by the flux at $4150$ \AA\,$<\lambda<4350$ \AA) around the [O{\sc ii}] doublet in bins of stellar mass (left) and in bins of local density (right). This figure shows the comparison between different stacks (for individual inspection we refer to Figure \ref{fig:stack_Mass_full}). We see a strong dependence of the [O{\sc ii}] strength on the stellar mass with higher stellar mass galaxies having weaker [O{\sc ii}] emission, as expected since most quenched galaxies are found at higher stellar masses and should have little to no emission. We also find a dependence of the [O{\sc ii}] strength on the local density with high density regions having  galaxies with weaker [O{\sc ii}] emission, again with massive quiescent galaxies dominating at higher densities likely the cause of this effect.}
\label{fig:stack_both}
\end{figure*}

\begin{figure}
\centering
\includegraphics[width=\linewidth]{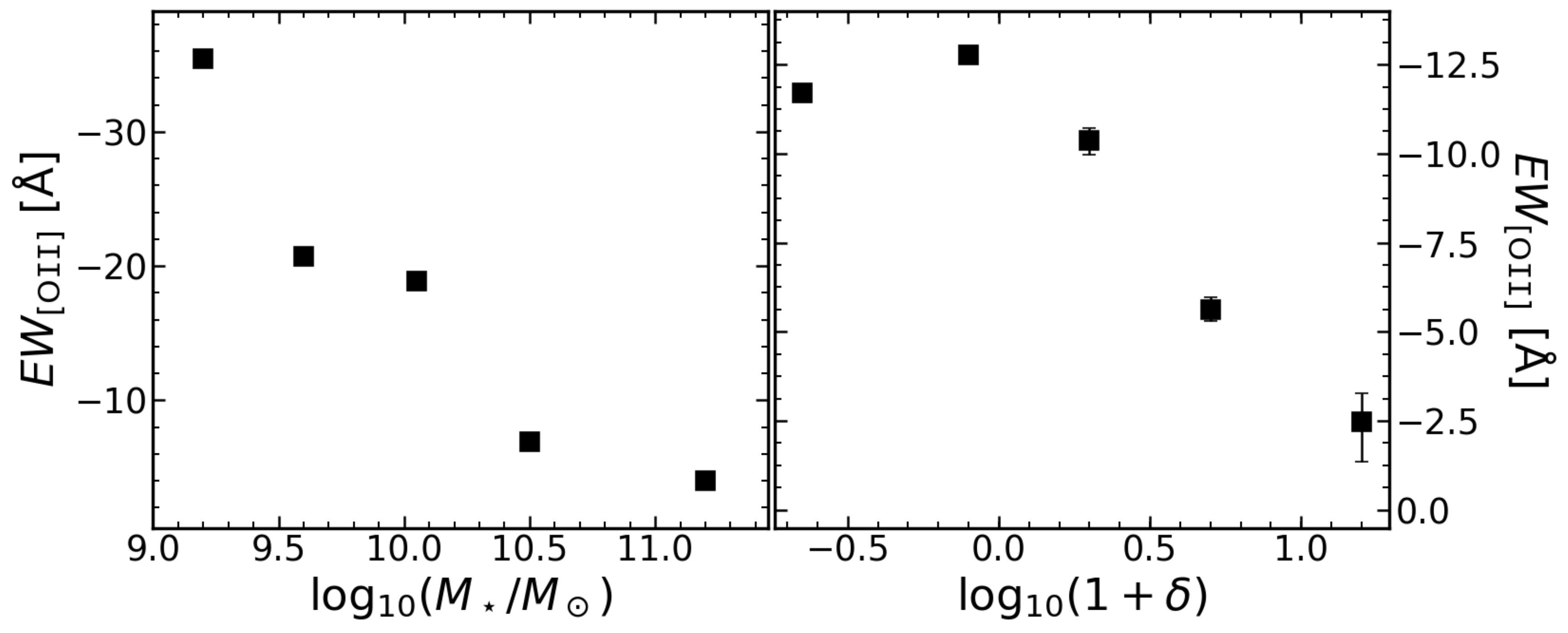}
\caption{\add{Line equivalent width for the spectral lines in [O{\sc ii}] as a function of the stellar mass range (left) and of the local density range (right) for the stacked spectra. We apply no dust correction to the stacked derived values since we assume spatial coincidence between the continuum and line emitting regions, and they are affected by dust in a similar manner.}}
\label{fig:stack_props}
\end{figure}

	\subsection{[O{\sc ii}] luminosity dependence on local overdensity}\label{ssection:oii}

We show in the lower panel of Figure \ref{fig:SFR_Mass} the resulting distribution of dust corrected [O{\sc ii}] luminosity for the sample  at $0.8<z<0.9$. The bulk of the population has $L_\mathrm{[OII]}\sim10^{41.5}\mathrm{erg\ s^{-1}}$ with the brightest in our sample reaching luminosities of $L_\mathrm{[OII]}\sim10^{43}\mathrm{erg\ s^{-1}}$.

When looking at galaxies in high- and low-density environments we find no significant difference in the median (excluding upper limits) dust-corrected [O{\sc ii}] luminosity at all stellar masses probed in our study. If we assume that the luminosity of the [O{\sc ii}] emission doublet is correlated with the galaxy SFR \citep[see e.g][]{kennicutt1998,kewley2004,darvish2015}, our results on [O{\sc ii}] show that their median luminosity (which traces star formation) is not affected by higher density environments. This is different than what we show when using SED fitting derived SFRs. We attribute this discrepancy to the fact that with our observational setup we measure [O{\sc ii}] luminosities more easily for star-forming galaxies than for quiescent galaxies (which are mostly upper limits). The SED fitting results do not suffer from the same problem, meaning that what is likely causing the differences is the quenched fraction as a function of density. Having higher fraction of quenched galaxies at high densities (see e.g. Figure \ref{fig:SFR_Quench}) will result in a lower median SFR value than what we would get from [O{\sc ii}] luminosities because we miss a fraction of that population (upper limits only in Figure \ref{fig:SFR_Mass}). If one includes the upper limits in the median calculation, we get qualitatively the same trends as we find for SED derived SFRs.

\add{We note, however, that differences may also arise if the [O{\sc ii}] emission is originating from other sources than star formation \citep[e.g. AGN, LINERs; see e.g.][]{yan2006,kocevski2011}, but we expect this to be a secondary effect due to the lower overall fraction of this type of objects \citep[e.g.][]{pentericci2013,ehlert2014,oh2014}.}

	\subsection{[O{\sc ii}] properties in stacked spectra}

We show in the left panel of Figure \ref{fig:stack_both} (see also Figure \ref{fig:stack_Mass_full} for individualized panels) the resulting spectra after stacking all galaxies in bins of stellar mass. We observe a strong decrease in [O{\sc ii}] line strength from low to high stellar masses (about a factor of $\sim10$ in flux from the lowest to the highest stellar mass bin). We also see the relative strength of the two doublet lines is changing with stellar mass. At lower masses the [O{\sc ii}]$\lambda$3729/[O{\sc ii}]$\lambda$3726 ratio is higher and seems to constantly decrease as we move towards higher masses. \add{This ratio is indicative of the electron density in the interstellar medium \citep[e.g.][]{seaton1957, canto1980, pradhan2006, darvish2015, sanders2016, kaasinen2017} and will be investigated in a subsequent paper.}

In Figure \ref{fig:stack_both} (right panel, see also Figure \ref{fig:stack_delta_full} for individualized panels), we show our findings of the stacked spectra in bins of local density. In terms of the [O{\sc ii}] emission we find a decreasing line strength from low to high density regions. Interestingly, in the three lowest density bins the difference in [O{\sc ii}] emission strength is appreciably smaller when compared to the two highest density bins. This decrease at $\log_{10}(\delta)\sim0-0.5$ hints at a break in star formation around these local overdensity values \citep[see e.g.][]{darvish2016}.

To quantify the properties of each line we have performed a double Gaussian fitting to [O{\sc ii}] using Equation \ref{eq:guassian_fit}. Results of the equivalent widths and fluxes of the [O{\sc ii}] doublet are summarized in Table \ref{table:line_props} and Figure \ref{fig:stack_props}. The qualitative remarks we made on the appearance of the spectral stacks are confirmed by our results after fitting each component.

We find a strong decrease in [O{\sc ii}] strength and line equivalent width with stellar mass (see Figure \ref{fig:stack_props}) with a factor of $\sim10$ between the lowest stellar mass bin ($9.0<\log_{10}\left(M_\star/\mathrm{M_\odot}\right)<9.4$) and the highest stellar mass bin ($\log_{10}\left(M_\star/\mathrm{M_\odot}\right)>10.7$) \citep[similar to results by e.g.][]{darvish2015,khostovan2016}. Performing the same analysis on the stacked spectra per local density bin, the [O{\sc ii}] line strength and equivalent width show broken relation with a "break" at $\log_{10}(1+\delta)\sim0.0-0.5$ that translates into a steeper relation at higher densities.


\section{Discussion}\label{section:discussion}

The survey that we present in this paper is able to select galaxies through their continuum emission and absorption features down to $\log_{10}\left(M_\star/\mathrm{M_\odot}\right)\sim10$ \add{and able to detect [O{\sc ii}] down to ${\sim5\times10^{-18}\ \mathrm{erg\, s^{-1}cm^{-2}}}$}. Since our sample is based on accurate measurements of redshifts, it is natural that it selects galaxies at lower stellar masses only if they have clear emission lines characteristic of star-forming galaxies. This means that our results on global trends with stellar mass below our completeness limit is biased against low star formation and passive galaxies (see e.g. Figure \ref{fig:mass_sfr_dist}). This fact alone is able to explain an apparent lack of trends in star formation related quantities (SFR and $ L_\mathrm{[OII]}$) at stellar masses below $\log_{10}\left(M_\star/\mathrm{M_\odot}\right)=10$, where we see no dependence whatsoever on local overdensity. In summary, our results for the lowest stellar mass bins (less than $10^{10}\mathrm{M_\odot}$) are likely based only on the star-forming population.

One important aspect to consider when looking for environmental effects on galaxy evolution is to attempt to distinguish between stellar mass driven and density driven mechanisms \citep[e.g.][]{peng2010b,kovac2014,darvish2016}. We attempt to address these issues by computing average quantities in different environments as a function of stellar mass (or at different stellar masses as a function of environment).  

Considering our results on galaxies with stellar masses above our completeness limit, we find small influence of environment on galaxy SFRs (from SED fitting) and $ L_\mathrm{[OII]}$. In higher density regions galaxies are typically less star-forming (Figure \ref{fig:SFR_Mass}, except at the highest stellar masses ($\log_{10}\left(M_\star/\mathrm{M_\odot}\right)>11$) but with comparable [O{\sc ii}] emission. This can be easily explained by the increase of the fraction of quenched galaxies in higher density regions. These trends support the scenario where environment plays a role in increasing the quiescent fraction of intermediate stellar mass galaxies at these redshifts ($z\approx0.84$). This fits well in the scenario where galaxies already have their star formation suppressed due to environmental effects as early as 7 billion years ago. At higher stellar masses, we see no differences in the average SFR and quiescent fractions, hinting that mass quenching should be effective enough to halt star formation even in low density regions \citep[see e.g][]{peng2010b} although it is not clear that environmental and stellar mass quenching are fully separable \citep[see e.g.][]{lee2015,darvish2016,kawinwanichakij2017}. This differential effect with stellar mass is a potential indicator that environment acts as a catalyst for star formation quenching in the sense that we are more likely to see galaxies quench at lower stellar masses if they reside in high density environments. We stress that for $10<\log_{10}\left(M_\star/\mathrm{M_\odot}\right)<10.25$ we find no differences in the median SFRs between low- and high-density regions and that is likely caused by a lack of quiescent galaxies close to our completeness limit that drives up the median value of the SFR for that bin. Since we see a rise in the quiescent fraction towards high-density regions on the lowest stellar mass bin we probe (see Figure \ref{fig:SFR_Quench} and also Appendix \ref{sec:completeness}), it is plausible that this is the reason for the observed results in this stellar mass bin.
 
Our findings corroborate those reported by \citet{sobral2011} which probed the same region using H$\alpha$ emitters. They are also consistent with others in the literature which already report a decrease in the star-forming fraction with projected galaxy density at similar redshifts \citep[e.g.][]{patel2009,muzzin2012}. \add{We also see similar trends of star formation with environment in lower redshift surveys \citep[e.g.][]{balogh2002,delpino2017}. This means that environmental effects are shaping the star formation in individual galaxies in a similar manner in the past 6 Gyr.} These effects are readily explained by the number of physical mechanisms (e.g. ram pressure stripping, tidal interactions) capable of stripping gas from galaxies and shut down any new star formation activity. While we note that these can in fact explain the observed trends in SFR with stellar mass and environment it is out of the scope of this paper to pinpoint which mechanisms are responsible for what we observe.

	\subsection{Halting of star formation in the outskirts of clusters}

Overall we find that the average SFR is lower in high-density regions, confirming what was reported by \citet{sobral2011} when studying H$\alpha$ emitters on the same structure. We report one order of magnitude difference in the average SFR from the lowest to the highest density region ($\sim10$ to $\sim1$ $\mathrm{M_\odot yr^{-1}}$). This trend with environment gives strength to the argument of environmentally driven quenching occurring within our superstructure. These signs of environmental quenching of star formation \citep[also seen in e.g.][]{patel2009,sobral2011,muzzin2012,santos2013} are distinct from what one structures that show a flat or reverse SFR-density relation \citep[e.g.][]{elbaz2007,ideue2009,tran2010,santos2014}.

One interesting result is the "break" that we find on the relation between [O{\sc ii}] line equivalent width and local overdensity that occurs at intermediate densities \citep[$\log_{10}(\delta)\sim0.0-0.5$, see Fig. \ref{fig:stack_props} and also e.g.][]{darvish2016}. We hypothesize that this corresponds to a typical density where environment quenching mechanisms are the most effective. The transition at $\log_{10}(\delta)\sim0.0-0.5$ is consistent with regions of filament-like densities (see transition from filament dominated to cluster dominated galaxies in Figure \ref{fig:dens_dist}). This result is compatible with intermediate density regions likely being the place of enhanced chances for galaxy encounters, promoting galaxy harassment related quenching mechanisms \citep[e.g.][]{moss2006,perez2009,li2009,tonnesen2012,darvish2014,malavasi2017}. It might also be caused by strong cluster-cluster interactions that are found to enhance star formation as well \citep[e.g.][]{stroe2014,stroe2015a}. To further reinforce the existence of such "break", we find that the fraction of quenched galaxies at intermediate stellar masses ($10<\log_{10}\left(M_\star/\mathrm{M_\odot}\right)<10.75$) increases by a factor of 2 at the same transition density, being roughly constant below and above. Galaxies at higher stellar masses are likely already quenched due to their own mass \citep[see e.g][]{peng2010b} and they are likely not much affected by the environment they are in.


\section{Conclusions}\label{section:conclusions}

We have presented in this paper an overview of the VIS$^3$COS survey, which targets a superstructure at $z\sim0.84$ with VIMOS/VLT high resolution spectra. \add{We report on trends with environment and stellar mass of the SFR and [O{\sc ii}] luminosity}. Our main findings are summarized as follows:

\begin{itemize}
\setlength\itemsep{1em}
\item Above our stellar mass completeness limit ($10^{10}\mathrm{M_\odot}$), galaxies in higher density regions have lower star formation rates at intermediate masses ($10<\log_{10}\left(M_\star/\mathrm{M_\odot}\right)<10.75$). At the highest masses (above $10^{10.75}\mathrm{M_\odot}$), the star formation activity is similar in low- and high-density environments indicating that mass quenching is probably dominant at high stellar masses.
\item We find that the fraction of quenched galaxies ($f_Q$) increases from $\sim10$\% to $\sim40-60$\% with increasing galaxy overdensity, but only for intermediate stellar mass galaxies ($10<\log_{10}\left(M_\star/\mathrm{M_\odot}\right)<10.75$). The most massive galaxies in our sample (above $10^{10.75}\mathrm{M_\odot}$) have a similar value of $f_Q\sim30-40$\% at all densities.
\item We find a break in [O{\sc ii}] strength and equivalent width in the stacked spectra in filament-like regions ($\log_{10}(\delta)\sim0.0-0.5$). We hypothesize that at these densities quenching mechanisms due to environment play an important role. This is consistent with the increase in the quenched galaxy fraction that we find for intermediate stellar mass galaxies.
\end{itemize}

\add{In summary, the results of this paper shed some light on the properties of galaxies in and around a superstructure on the COSMOS field. In this paper we have focused on the overall properties of the sample in our survey and the general trends that we find on galaxy properties with respect to environment. More detailed studies focusing on the individual star formation activity of galaxies, galaxy morphology, and electron density estimates will be discussed in forthcoming papers.}


\section*{Acknowledgements}

We thank the anonymous referee for making valuable suggestions that helped to improve this paper. This work was supported by Funda\c{c}\~{a}o para a Ci\^{e}ncia e a Tecnologia (FCT) through the research grant UID/FIS/04434/2013. APA, PhD::SPACE fellow, acknowledges support from the FCT through the fellowship PD/BD/52706/2014. DS acknowledges financial support from the Netherlands Organisation for Scientific research (NWO) through a Veni fellowship and from Lancaster University through an Early Career Internal Grant A100679. BD acknowledges financial support from NASA through the Astrophysics Data Analysis Program (ADAP), grant number NNX12AE20G, and the National Science Foundation, grant number 1716907. PNB is grateful for support from STFC via grant ST/M001229/1.

This work was only possible by the use of the following \textsc{python} packages: NumPy \& SciPy \citep{walt2011,jones2001}, Matplotlib \citep{hunter2007}, and Astropy \citep{robitaille2013}.


\bibliographystyle{aa}
\bibliography{refs}

\begin{thebibliography}{138}
\expandafter\ifx\csname natexlab\endcsname\relax\def\natexlab#1{#1}\fi

\bibitem[{{Arag{\'o}n-Calvo} {et~al.}(2010){Arag{\'o}n-Calvo}, {van de
  Weygaert}, \& {Jones}}]{aragon-calvo2010}
{Arag{\'o}n-Calvo}, M.~A., {van de Weygaert}, R., \& {Jones}, B.~J.~T. 2010,
  \mnras, 408, 2163

\bibitem[{{Astropy Collaboration} {et~al.}(2013){Astropy Collaboration},
  {Robitaille}, {Tollerud}, {Greenfield}, {Droettboom}, {Bray}, {Aldcroft},
  {Davis}, {Ginsburg}, {Price-Whelan}, {Kerzendorf}, {Conley}, {Crighton},
  {Barbary}, {Muna}, {Ferguson}, {Grollier}, {Parikh}, {Nair}, {Unther},
  {Deil}, {Woillez}, {Conseil}, {Kramer}, {Turner}, {Singer}, {Fox}, {Weaver},
  {Zabalza}, {Edwards}, {Azalee Bostroem}, {Burke}, {Casey}, {Crawford},
  {Dencheva}, {Ely}, {Jenness}, {Labrie}, {Lim}, {Pierfederici}, {Pontzen},
  {Ptak}, {Refsdal}, {Servillat}, \& {Streicher}}]{robitaille2013}
{Astropy Collaboration}, {Robitaille}, T.~P., {Tollerud}, E.~J., {et~al.} 2013,
  \aap, 558, A33

\bibitem[{{Baldry} {et~al.}(2006){Baldry}, {Balogh}, {Bower}, {Glazebrook},
  {Nichol}, {Bamford}, \& {Budavari}}]{baldry2006}
{Baldry}, I.~K., {Balogh}, M.~L., {Bower}, R.~G., {et~al.} 2006, \mnras, 373,
  469

\bibitem[{{Balogh} {et~al.}(2004){Balogh}, {Eke}, {Miller}, {Lewis}, {Bower},
  {Couch}, {Nichol}, {Bland-Hawthorn}, {Baldry}, {Baugh}, {Bridges}, {Cannon},
  {Cole}, {Colless}, {Collins}, {Cross}, {Dalton}, {de Propris}, {Driver},
  {Efstathiou}, {Ellis}, {Frenk}, {Glazebrook}, {Gomez}, {Gray}, {Hawkins},
  {Jackson}, {Lahav}, {Lumsden}, {Maddox}, {Madgwick}, {Norberg}, {Peacock},
  {Percival}, {Peterson}, {Sutherland}, \& {Taylor}}]{balogh2004}
{Balogh}, M., {Eke}, V., {Miller}, C., {et~al.} 2004, \mnras, 348, 1355

\bibitem[{{Balogh} {et~al.}(2002){Balogh}, {Couch}, {Smail}, {Bower}, \&
  {Glazebrook}}]{balogh2002}
{Balogh}, M.~L., {Couch}, W.~J., {Smail}, I., {Bower}, R.~G., \& {Glazebrook},
  K. 2002, \mnras, 335, 10

\bibitem[{{Balogh} {et~al.}(2014){Balogh}, {McGee}, {Mok}, {Wilman},
  {Finoguenov}, {Bower}, {Mulchaey}, {Parker}, \& {Tanaka}}]{balogh2014}
{Balogh}, M.~L., {McGee}, S.~L., {Mok}, A., {et~al.} 2014, \mnras, 443, 2679

\bibitem[{{Bamford} {et~al.}(2009){Bamford}, {Nichol}, {Baldry}, {Land},
  {Lintott}, {Schawinski}, {Slosar}, {Szalay}, {Thomas}, {Torki}, {Andreescu},
  {Edmondson}, {Miller}, {Murray}, {Raddick}, \& {Vandenberg}}]{bamford2009}
{Bamford}, S.~P., {Nichol}, R.~C., {Baldry}, I.~K., {et~al.} 2009, \mnras, 393,
  1324

\bibitem[{{Beers} {et~al.}(1990){Beers}, {Flynn}, \& {Gebhardt}}]{beers1990}
{Beers}, T.~C., {Flynn}, K., \& {Gebhardt}, K. 1990, \aj, 100, 32

\bibitem[{{Bekki}(2009)}]{bekki2009}
{Bekki}, K. 2009, \mnras, 399, 2221

\bibitem[{{Best}(2004)}]{best2004}
{Best}, P.~N. 2004, \mnras, 351, 70

\bibitem[{{Boselli} {et~al.}(2014){Boselli}, {Cortese}, {Boquien}, {Boissier},
  {Catinella}, {Gavazzi}, {Lagos}, \& {Saintonge}}]{boselli2014b}
{Boselli}, A., {Cortese}, L., {Boquien}, M., {et~al.} 2014, \aap, 564, A67

\bibitem[{{Brown} {et~al.}(2017){Brown}, {Catinella}, {Cortese}, {Lagos},
  {Dav{\'e}}, {Kilborn}, {Haynes}, {Giovanelli}, \&
  {Rafieferantsoa}}]{brown2017}
{Brown}, T., {Catinella}, B., {Cortese}, L., {et~al.} 2017, \mnras, 466, 1275

\bibitem[{{Bruzual} \& {Charlot}(2003)}]{bruzual2003}
{Bruzual}, G. \& {Charlot}, S. 2003, \mnras, 344, 1000

\bibitem[{{Burgarella} {et~al.}(2013){Burgarella}, {Buat}, {Gruppioni},
  {Cucciati}, {Heinis}, {Berta}, {B{\'e}thermin}, {Bock}, {Cooray}, {Dunlop},
  {Farrah}, {Franceschini}, {Le Floc'h}, {Lutz}, {Magnelli}, {Nordon},
  {Oliver}, {Page}, {Popesso}, {Pozzi}, {Riguccini}, {Vaccari}, \&
  {Viero}}]{burgarella2013}
{Burgarella}, D., {Buat}, V., {Gruppioni}, C., {et~al.} 2013, \aap, 554, A70

\bibitem[{{Canto} {et~al.}(1980){Canto}, {Meaburn}, {Theokas}, \&
  {Elliott}}]{canto1980}
{Canto}, J., {Meaburn}, J., {Theokas}, A.~C., \& {Elliott}, K.~H. 1980, \mnras,
  193, 911

\bibitem[{{Carollo} {et~al.}(2013){Carollo}, {Bschorr}, {Renzini}, {Lilly},
  {Capak}, {Cibinel}, {Ilbert}, {Onodera}, {Scoville}, {Cameron}, {Mobasher},
  {Sanders}, \& {Taniguchi}}]{carollo2013}
{Carollo}, C.~M., {Bschorr}, T.~J., {Renzini}, A., {et~al.} 2013, \apj, 773,
  112

\bibitem[{{Chabrier}(2003)}]{chabrier2003}
{Chabrier}, G. 2003, \apjl, 586, L133

\bibitem[{{Charlot} \& {Fall}(2000)}]{charlot2000}
{Charlot}, S. \& {Fall}, S.~M. 2000, \apj, 539, 718

\bibitem[{{Cohen} {et~al.}(2017){Cohen}, {Hickox}, {Wegner}, {Einasto}, \&
  {Vennik}}]{cohen2017}
{Cohen}, S.~A., {Hickox}, R.~C., {Wegner}, G.~A., {Einasto}, M., \& {Vennik},
  J. 2017, \apj, 835, 56

\bibitem[{{Cooper} {et~al.}(2008){Cooper}, {Newman}, {Weiner}, {Yan},
  {Willmer}, {Bundy}, {Coil}, {Conselice}, {Davis}, {Faber}, {Gerke},
  {Guhathakurta}, {Koo}, \& {Noeske}}]{cooper2008}
{Cooper}, M.~C., {Newman}, J.~A., {Weiner}, B.~J., {et~al.} 2008, \mnras, 383,
  1058

\bibitem[{{Cortese} {et~al.}(2010){Cortese}, {Davies}, {Pohlen}, {Baes},
  {Bendo}, {Bianchi}, {Boselli}, {De Looze}, {Fritz}, {Verstappen}, {Bomans},
  {Clemens}, {Corbelli}, {Dariush}, {di Serego Alighieri}, {Fadda},
  {Garcia-Appadoo}, {Gavazzi}, {Giovanardi}, {Grossi}, {Hughes}, {Hunt},
  {Jones}, {Madden}, {Pierini}, {Sabatini}, {Smith}, {Vlahakis}, {Xilouris}, \&
  {Zibetti}}]{cortese2010}
{Cortese}, L., {Davies}, J.~I., {Pohlen}, M., {et~al.} 2010, \aap, 518, L49

\bibitem[{{Crain} {et~al.}(2015){Crain}, {Schaye}, {Bower}, {Furlong},
  {Schaller}, {Theuns}, {Dalla Vecchia}, {Frenk}, {McCarthy}, {Helly},
  {Jenkins}, {Rosas-Guevara}, {White}, \& {Trayford}}]{crain2015}
{Crain}, R.~A., {Schaye}, J., {Bower}, R.~G., {et~al.} 2015, \mnras, 450, 1937

\bibitem[{{Cucciati} {et~al.}(2017){Cucciati}, {Davidzon}, {Bolzonella},
  {Granett}, {De Lucia}, {Branchini}, {Zamorani}, {Iovino}, {Garilli}, {Guzzo},
  {Scodeggio}, {de la Torre}, {Abbas}, {Adami}, {Arnouts}, {Bottini}, {Cappi},
  {Franzetti}, {Fritz}, {Krywult}, {Le Brun}, {Le F{\`e}vre}, {Maccagni},
  {Ma{\l}ek}, {Marulli}, {Moutard}, {Polletta}, {Pollo}, {Tasca}, {Tojeiro},
  {Vergani}, {Zanichelli}, {Bel}, {Blaizot}, {Coupon}, {Hawken}, {Ilbert},
  {Moscardini}, {Peacock}, \& {Gargiulo}}]{cucciati2017}
{Cucciati}, O., {Davidzon}, I., {Bolzonella}, M., {et~al.} 2017, \aap, 602, A15

\bibitem[{{Cucciati} {et~al.}(2010{\natexlab{a}}){Cucciati}, {Iovino}, {Kova{\v
  c}}, {Scodeggio}, {Lilly}, {Bolzonella}, {Bardelli}, {Vergani}, {Tasca},
  {Zucca}, {Zamorani}, {Pozzetti}, {Knobel}, {Oesch}, {Lamareille}, {Caputi},
  {Kampczyk}, {Tresse}, {Maier}, {Carollo}, {Contini}, {Kneib}, {Le F{\`e}vre},
  {Mainieri}, {Renzini}, {Bongiorno}, {Coppa}, {de la Torre}, {de Ravel},
  {Franzetti}, {Garilli}, {Le Borgne}, {Le Brun}, {Mignoli}, {Pell{\`o}},
  {Peng}, {Perez-Montero}, {Ricciardelli}, {Silverman}, {Tanaka}, {Koekemoer},
  {Scoville}, {Abbas}, {Bottini}, {Cappi}, {Cassata}, {Cimatti}, {Guzzo},
  {Leauthaud}, {Maccagni}, {Marinoni}, {McCracken}, {Memeo}, {Meneux},
  {Porciani}, \& {Scaramella}}]{cucciati2010b}
{Cucciati}, O., {Iovino}, A., {Kova{\v c}}, K., {et~al.} 2010{\natexlab{a}},
  \aap, 524, A2

\bibitem[{{Cucciati} {et~al.}(2010{\natexlab{b}}){Cucciati}, {Marinoni},
  {Iovino}, {Bardelli}, {Adami}, {Mazure}, {Scodeggio}, {Maccagni}, {Temporin},
  {Zucca}, {De Lucia}, {Blaizot}, {Garilli}, {Meneux}, {Zamorani}, {Le
  F{\`e}vre}, {Cappi}, {Guzzo}, {Bottini}, {Le Brun}, {Tresse}, {Vettolani},
  {Zanichelli}, {Arnouts}, {Bolzonella}, {Charlot}, {Ciliegi}, {Contini},
  {Foucaud}, {Franzetti}, {Gavignaud}, {Ilbert}, {Lamareille}, {McCracken},
  {Marano}, {Merighi}, {Paltani}, {Pell{\`o}}, {Pollo}, {Pozzetti}, {Vergani},
  \& {P{\'e}rez-Montero}}]{cucciati2010a}
{Cucciati}, O., {Marinoni}, C., {Iovino}, A., {et~al.} 2010{\natexlab{b}},
  \aap, 520, A42

\bibitem[{{Cucciati} {et~al.}(2014){Cucciati}, {Zamorani}, {Lemaux},
  {Bardelli}, {Cimatti}, {Le F{\`e}vre}, {Cassata}, {Garilli}, {Le Brun},
  {Maccagni}, {Pentericci}, {Tasca}, {Thomas}, {Vanzella}, {Zucca}, {Amorin},
  {Capak}, {Cassar{\`a}}, {Castellano}, {Cuby}, {de la Torre}, {Durkalec},
  {Fontana}, {Giavalisco}, {Grazian}, {Hathi}, {Ilbert}, {Moreau}, {Paltani},
  {Ribeiro}, {Salvato}, {Schaerer}, {Scodeggio}, {Sommariva}, {Talia},
  {Taniguchi}, {Tresse}, {Vergani}, {Wang}, {Charlot}, {Contini}, {Fotopoulou},
  {L{\'o}pez-Sanjuan}, {Mellier}, \& {Scoville}}]{cucciati2014}
{Cucciati}, O., {Zamorani}, G., {Lemaux}, B.~C., {et~al.} 2014, \aap, 570, A16

\bibitem[{{da Cunha} {et~al.}(2008){da Cunha}, {Charlot}, \&
  {Elbaz}}]{cunha2008}
{da Cunha}, E., {Charlot}, S., \& {Elbaz}, D. 2008, \mnras, 388, 1595

\bibitem[{{Darvish} {et~al.}(2017){Darvish}, {Mobasher}, {Martin}, {Sobral},
  {Scoville}, {Stroe}, {Hemmati}, \& {Kartaltepe}}]{darvish2017}
{Darvish}, B., {Mobasher}, B., {Martin}, D.~C., {et~al.} 2017, \apj, 837, 16

\bibitem[{{Darvish} {et~al.}(2015{\natexlab{a}}){Darvish}, {Mobasher},
  {Sobral}, {Hemmati}, {Nayyeri}, \& {Shivaei}}]{darvish2015}
{Darvish}, B., {Mobasher}, B., {Sobral}, D., {et~al.} 2015{\natexlab{a}}, \apj,
  814, 84

\bibitem[{{Darvish} {et~al.}(2016){Darvish}, {Mobasher}, {Sobral}, {Rettura},
  {Scoville}, {Faisst}, \& {Capak}}]{darvish2016}
{Darvish}, B., {Mobasher}, B., {Sobral}, D., {et~al.} 2016, \apj, 825, 113

\bibitem[{{Darvish} {et~al.}(2015{\natexlab{b}}){Darvish}, {Mobasher},
  {Sobral}, {Scoville}, \& {Aragon-Calvo}}]{darvish2015a}
{Darvish}, B., {Mobasher}, B., {Sobral}, D., {Scoville}, N., \& {Aragon-Calvo},
  M. 2015{\natexlab{b}}, \apj, 805, 121

\bibitem[{{Darvish} {et~al.}(2014){Darvish}, {Sobral}, {Mobasher}, {Scoville},
  {Best}, {Sales}, \& {Smail}}]{darvish2014}
{Darvish}, B., {Sobral}, D., {Mobasher}, B., {et~al.} 2014, \apj, 796, 51

\bibitem[{{Draine}(2004)}]{draine2004}
{Draine}, B.~T. 2004, in The Cold Universe, Saas-Fee Advanced Course 32,
  Springer-Verlag, 308 pages, 129 figures, Lecture Notes 2002 of the Swiss
  Society for Astronomy and Astrophysics (SSAA), Springer, 2004. Edited by A.W.
  Blain, F. Combes, B.T. Draine, D. Pfenniger and Y. Revaz, ISBN 354040838x, p.
  213, ed. A.~W. {Blain}, F.~{Combes}, B.~T. {Draine}, D.~{Pfenniger}, \&
  Y.~{Revaz}, 213

\bibitem[{{Dressler}(1980)}]{dressler1980}
{Dressler}, A. 1980, \apj, 236, 351

\bibitem[{{Dressler}(1984)}]{dressler1984}
{Dressler}, A. 1984, \araa, 22, 185

\bibitem[{{Ehlert} {et~al.}(2014){Ehlert}, {von der Linden}, {Allen}, {Brandt},
  {Xue}, {Luo}, {Mantz}, {Morris}, {Applegate}, \& {Kelly}}]{ehlert2014}
{Ehlert}, S., {von der Linden}, A., {Allen}, S.~W., {et~al.} 2014, \mnras, 437,
  1942

\bibitem[{{Elbaz} {et~al.}(2007){Elbaz}, {Daddi}, {Le Borgne}, {Dickinson},
  {Alexander}, {Chary}, {Starck}, {Brandt}, {Kitzbichler}, {MacDonald},
  {Nonino}, {Popesso}, {Stern}, \& {Vanzella}}]{elbaz2007}
{Elbaz}, D., {Daddi}, E., {Le Borgne}, D., {et~al.} 2007, \aap, 468, 33

\bibitem[{{Ellison} {et~al.}(2008){Ellison}, {Patton}, {Simard}, \&
  {McConnachie}}]{ellison2008}
{Ellison}, S.~L., {Patton}, D.~R., {Simard}, L., \& {McConnachie}, A.~W. 2008,
  \aj, 135, 1877

\bibitem[{{Finoguenov} {et~al.}(2007){Finoguenov}, {Guzzo}, {Hasinger},
  {Scoville}, {Aussel}, {B{\"o}hringer}, {Brusa}, {Capak}, {Cappelluti},
  {Comastri}, {Giodini}, {Griffiths}, {Impey}, {Koekemoer}, {Kneib},
  {Leauthaud}, {Le F{\`e}vre}, {Lilly}, {Mainieri}, {Massey}, {McCracken},
  {Mobasher}, {Murayama}, {Peacock}, {Sakelliou}, {Schinnerer}, {Silverman},
  {Smol{\v c}i{\'c}}, {Taniguchi}, {Tasca}, {Taylor}, {Trump}, \&
  {Zamorani}}]{finoguenov2007}
{Finoguenov}, A., {Guzzo}, L., {Hasinger}, G., {et~al.} 2007, \apjs, 172, 182

\bibitem[{{Fumagalli} {et~al.}(2014){Fumagalli}, {Labb{\'e}}, {Patel}, {Franx},
  {van Dokkum}, {Brammer}, {da Cunha}, {F{\"o}rster Schreiber}, {Kriek},
  {Quadri}, {Rix}, {Wake}, {Whitaker}, {Lundgren}, {Marchesini}, {Maseda},
  {Momcheva}, {Nelson}, {Pacifici}, \& {Skelton}}]{fumagalli2014}
{Fumagalli}, M., {Labb{\'e}}, I., {Patel}, S.~G., {et~al.} 2014, \apj, 796, 35

\bibitem[{{Gallazzi} {et~al.}(2009){Gallazzi}, {Bell}, {Wolf}, {Gray},
  {Papovich}, {Barden}, {Peng}, {Meisenheimer}, {Heymans}, {van Kampen},
  {Gilmour}, {Balogh}, {McIntosh}, {Bacon}, {Barazza}, {B{\"o}hm}, {Caldwell},
  {H{\"a}u{\ss}ler}, {Jahnke}, {Jogee}, {Lane}, {Robaina}, {Sanchez}, {Taylor},
  {Wisotzki}, \& {Zheng}}]{gallazzi2009}
{Gallazzi}, A., {Bell}, E.~F., {Wolf}, C., {et~al.} 2009, \apj, 690, 1883

\bibitem[{{Genel} {et~al.}(2014){Genel}, {Vogelsberger}, {Springel}, {Sijacki},
  {Nelson}, {Snyder}, {Rodriguez-Gomez}, {Torrey}, \& {Hernquist}}]{genel2014}
{Genel}, S., {Vogelsberger}, M., {Springel}, V., {et~al.} 2014, \mnras, 445,
  175

\bibitem[{{Giovanelli} \& {Haynes}(1985)}]{giovanelli1985}
{Giovanelli}, R. \& {Haynes}, M.~P. 1985, \apj, 292, 404

\bibitem[{{G{\'o}mez} {et~al.}(2003){G{\'o}mez}, {Nichol}, {Miller}, {Balogh},
  {Goto}, {Zabludoff}, {Romer}, {Bernardi}, {Sheth}, {Hopkins}, {Castander},
  {Connolly}, {Schneider}, {Brinkmann}, {Lamb}, {SubbaRao}, \&
  {York}}]{gomez2003}
{G{\'o}mez}, P.~L., {Nichol}, R.~C., {Miller}, C.~J., {et~al.} 2003, \apj, 584,
  210

\bibitem[{{Hayashi} {et~al.}(2014){Hayashi}, {Kodama}, {Koyama}, {Tadaki},
  {Tanaka}, {Shimakawa}, {Matsuda}, {Sobral}, {Best}, \& {Smail}}]{hayashi2014}
{Hayashi}, M., {Kodama}, T., {Koyama}, Y., {et~al.} 2014, \mnras, 439, 2571

\bibitem[{{Henriques} {et~al.}(2015){Henriques}, {White}, {Thomas}, {Angulo},
  {Guo}, {Lemson}, {Springel}, \& {Overzier}}]{henriques2015}
{Henriques}, B.~M.~B., {White}, S.~D.~M., {Thomas}, P.~A., {et~al.} 2015,
  \mnras, 451, 2663

\bibitem[{{Hogg} {et~al.}(2004){Hogg}, {Blanton}, {Brinchmann}, {Eisenstein},
  {Schlegel}, {Gunn}, {McKay}, {Rix}, {Bahcall}, {Brinkmann}, \&
  {Meiksin}}]{hogg2004}
{Hogg}, D.~W., {Blanton}, M.~R., {Brinchmann}, J., {et~al.} 2004, \apjl, 601,
  L29

\bibitem[{Hunter(2007)}]{hunter2007}
Hunter, J.~D. 2007, Computing In Science \& Engineering, 9, 90

\bibitem[{{Ideue} {et~al.}(2009){Ideue}, {Nagao}, {Taniguchi}, {Shioya},
  {Saito}, {Murayama}, {Sasaki}, {Trump}, {Koekemoer}, {Aussel}, {Capak},
  {Ilbert}, {McCracken}, {Mobasher}, {Salvato}, {Sanders}, \&
  {Scoville}}]{ideue2009}
{Ideue}, Y., {Nagao}, T., {Taniguchi}, Y., {et~al.} 2009, \apj, 700, 971

\bibitem[{{Ilbert} {et~al.}(2009){Ilbert}, {Capak}, {Salvato}, {Aussel},
  {McCracken}, {Sanders}, {Scoville}, {Kartaltepe}, {Arnouts}, {Le Floc'h},
  {Mobasher}, {Taniguchi}, {Lamareille}, {Leauthaud}, {Sasaki}, {Thompson},
  {Zamojski}, {Zamorani}, {Bardelli}, {Bolzonella}, {Bongiorno}, {Brusa},
  {Caputi}, {Carollo}, {Contini}, {Cook}, {Coppa}, {Cucciati}, {de la Torre},
  {de Ravel}, {Franzetti}, {Garilli}, {Hasinger}, {Iovino}, {Kampczyk},
  {Kneib}, {Knobel}, {Kovac}, {Le Borgne}, {Le Brun}, {F{\`e}vre}, {Lilly},
  {Looper}, {Maier}, {Mainieri}, {Mellier}, {Mignoli}, {Murayama}, {Pell{\`o}},
  {Peng}, {P{\'e}rez-Montero}, {Renzini}, {Ricciardelli}, {Schiminovich},
  {Scodeggio}, {Shioya}, {Silverman}, {Surace}, {Tanaka}, {Tasca}, {Tresse},
  {Vergani}, \& {Zucca}}]{ilbert2009}
{Ilbert}, O., {Capak}, P., {Salvato}, M., {et~al.} 2009, \apj, 690, 1236

\bibitem[{{Ilbert} {et~al.}(2013){Ilbert}, {McCracken}, {Le F{\`e}vre},
  {Capak}, {Dunlop}, {Karim}, {Renzini}, {Caputi}, {Boissier}, {Arnouts},
  {Aussel}, {Comparat}, {Guo}, {Hudelot}, {Kartaltepe}, {Kneib}, {Krogager},
  {Le Floc'h}, {Lilly}, {Mellier}, {Milvang-Jensen}, {Moutard}, {Onodera},
  {Richard}, {Salvato}, {Sanders}, {Scoville}, {Silverman}, {Taniguchi},
  {Tasca}, {Thomas}, {Toft}, {Tresse}, {Vergani}, {Wolk}, \&
  {Zirm}}]{ilbert2013}
{Ilbert}, O., {McCracken}, H.~J., {Le F{\`e}vre}, O., {et~al.} 2013, \aap, 556,
  A55

\bibitem[{{Ilbert} {et~al.}(2010){Ilbert}, {Salvato}, {Le Floc'h}, {Aussel},
  {Capak}, {McCracken}, {Mobasher}, {Kartaltepe}, {Scoville}, {Sanders},
  {Arnouts}, {Bundy}, {Cassata}, {Kneib}, {Koekemoer}, {Le F{\`e}vre}, {Lilly},
  {Surace}, {Taniguchi}, {Tasca}, {Thompson}, {Tresse}, {Zamojski}, {Zamorani},
  \& {Zucca}}]{ilbert2010}
{Ilbert}, O., {Salvato}, M., {Le Floc'h}, E., {et~al.} 2010, \apj, 709, 644

\bibitem[{{Iovino} {et~al.}(2010){Iovino}, {Cucciati}, {Scodeggio}, {Knobel},
  {Kova{\v c}}, {Lilly}, {Bolzonella}, {Tasca}, {Zamorani}, {Zucca}, {Caputi},
  {Pozzetti}, {Oesch}, {Lamareille}, {Halliday}, {Bardelli}, {Finoguenov},
  {Guzzo}, {Kampczyk}, {Maier}, {Tanaka}, {Vergani}, {Carollo}, {Contini},
  {Kneib}, {Le F{\`e}vre}, {Mainieri}, {Renzini}, {Bongiorno}, {Coppa}, {de la
  Torre}, {de Ravel}, {Franzetti}, {Garilli}, {Le Borgne}, {Le Brun},
  {Mignoli}, {Pell{\`o}}, {Peng}, {Perez-Montero}, {Ricciardelli}, {Silverman},
  {Tresse}, {Abbas}, {Bottini}, {Cappi}, {Cassata}, {Cimatti}, {Koekemoer},
  {Leauthaud}, {Maccagni}, {Marinoni}, {McCracken}, {Memeo}, {Meneux},
  {Porciani}, {Scaramella}, {Schiminovich}, \& {Scoville}}]{iovino2010}
{Iovino}, A., {Cucciati}, O., {Scodeggio}, M., {et~al.} 2010, \aap, 509, A40

\bibitem[{{Iovino} {et~al.}(2016){Iovino}, {Petropoulou}, {Scodeggio},
  {Bolzonella}, {Zamorani}, {Bardelli}, {Cucciati}, {Pozzetti}, {Tasca},
  {Vergani}, {Zucca}, {Finoguenov}, {Ilbert}, {Tanaka}, {Salvato}, {Kova{\v
  c}}, \& {Cassata}}]{iovino2016}
{Iovino}, A., {Petropoulou}, V., {Scodeggio}, M., {et~al.} 2016, \aap, 592, A78

\bibitem[{Jones {et~al.}(2001)Jones, Oliphant, Peterson, {et~al.}}]{jones2001}
Jones, E., Oliphant, T., Peterson, P., {et~al.} 2001, {SciPy}: Open source
  scientific tools for {Python}, [Online; accessed 2016-03-23]

\bibitem[{{Kaasinen} {et~al.}(2017){Kaasinen}, {Bian}, {Groves}, {Kewley}, \&
  {Gupta}}]{kaasinen2017}
{Kaasinen}, M., {Bian}, F., {Groves}, B., {Kewley}, L.~J., \& {Gupta}, A. 2017,
  \mnras, 465, 3220

\bibitem[{{Karim} {et~al.}(2011){Karim}, {Schinnerer},
  {Mart{\'{\i}}nez-Sansigre}, {Sargent}, {van der Wel}, {Rix}, {Ilbert},
  {Smol{\v c}i{\'c}}, {Carilli}, {Pannella}, {Koekemoer}, {Bell}, \&
  {Salvato}}]{karim2011}
{Karim}, A., {Schinnerer}, E., {Mart{\'{\i}}nez-Sansigre}, A., {et~al.} 2011,
  \apj, 730, 61

\bibitem[{{Kauffmann} {et~al.}(2004){Kauffmann}, {White}, {Heckman},
  {M{\'e}nard}, {Brinchmann}, {Charlot}, {Tremonti}, \&
  {Brinkmann}}]{kauffmann2004}
{Kauffmann}, G., {White}, S.~D.~M., {Heckman}, T.~M., {et~al.} 2004, \mnras,
  353, 713

\bibitem[{{Kawinwanichakij} {et~al.}(2017){Kawinwanichakij}, {Papovich},
  {Quadri}, {Glazebrook}, {Kacprzak}, {Allen}, {Bell}, {Croton}, {Dekel},
  {Ferguson}, {Forrest}, {Grogin}, {Guo}, {Kocevski}, {Koekemoer}, {Labb{\'e}},
  {Lucas}, {Nanayakkara}, {Spitler}, {Straatman}, {Tran}, {Tomczak}, \& {van
  Dokkum}}]{kawinwanichakij2017}
{Kawinwanichakij}, L., {Papovich}, C., {Quadri}, R.~F., {et~al.} 2017, \apj,
  847, 134

\bibitem[{{Kennicutt}(1998)}]{kennicutt1998}
{Kennicutt}, Jr., R.~C. 1998, \araa, 36, 189

\bibitem[{{Kewley} {et~al.}(2006){Kewley}, {Geller}, \& {Barton}}]{kewley2006}
{Kewley}, L.~J., {Geller}, M.~J., \& {Barton}, E.~J. 2006, \aj, 131, 2004

\bibitem[{{Kewley} {et~al.}(2004){Kewley}, {Geller}, \& {Jansen}}]{kewley2004}
{Kewley}, L.~J., {Geller}, M.~J., \& {Jansen}, R.~A. 2004, \aj, 127, 2002

\bibitem[{{Khostovan} {et~al.}(2015){Khostovan}, {Sobral}, {Mobasher}, {Best},
  {Smail}, {Stott}, {Hemmati}, \& {Nayyeri}}]{khostovan2015}
{Khostovan}, A.~A., {Sobral}, D., {Mobasher}, B., {et~al.} 2015, \mnras, 452,
  3948

\bibitem[{{Khostovan} {et~al.}(2016){Khostovan}, {Sobral}, {Mobasher}, {Smail},
  {Darvish}, {Nayyeri}, {Hemmati}, \& {Stott}}]{khostovan2016}
{Khostovan}, A.~A., {Sobral}, D., {Mobasher}, B., {et~al.} 2016, \mnras, 463,
  2363

\bibitem[{{Kocevski} {et~al.}(2011){Kocevski}, {Lemaux}, {Lubin}, {Shapley},
  {Gal}, \& {Squires}}]{kocevski2011}
{Kocevski}, D.~D., {Lemaux}, B.~C., {Lubin}, L.~M., {et~al.} 2011, \apjl, 737,
  L38

\bibitem[{{Kodama} {et~al.}(2004){Kodama}, {Balogh}, {Smail}, {Bower}, \&
  {Nakata}}]{kodama2004}
{Kodama}, T., {Balogh}, M.~L., {Smail}, I., {Bower}, R.~G., \& {Nakata}, F.
  2004, \mnras, 354, 1103

\bibitem[{{Koekemoer} {et~al.}(2007){Koekemoer}, {Aussel}, {Calzetti}, {Capak},
  {Giavalisco}, {Kneib}, {Leauthaud}, {Le F{\`e}vre}, {McCracken}, {Massey},
  {Mobasher}, {Rhodes}, {Scoville}, \& {Shopbell}}]{koekemoer2007}
{Koekemoer}, A.~M., {Aussel}, H., {Calzetti}, D., {et~al.} 2007, ApJS, 172, 196

\bibitem[{{Kova{\v c}} {et~al.}(2014){Kova{\v c}}, {Lilly}, {Knobel},
  {Bschorr}, {Peng}, {Carollo}, {Contini}, {Kneib}, {Le F{\'e}vre}, {Mainieri},
  {Renzini}, {Scodeggio}, {Zamorani}, {Bardelli}, {Bolzonella}, {Bongiorno},
  {Caputi}, {Cucciati}, {de la Torre}, {de Ravel}, {Franzetti}, {Garilli},
  {Iovino}, {Kampczyk}, {Lamareille}, {Le Borgne}, {Le Brun}, {Maier},
  {Mignoli}, {Oesch}, {Pello}, {Montero}, {Presotto}, {Silverman}, {Tanaka},
  {Tasca}, {Tresse}, {Vergani}, {Zucca}, {Aussel}, {Koekemoer}, {Le Floc'h},
  {Moresco}, \& {Pozzetti}}]{kovac2014}
{Kova{\v c}}, K., {Lilly}, S.~J., {Knobel}, C., {et~al.} 2014, \mnras, 438, 717

\bibitem[{{Koyama} {et~al.}(2017){Koyama}, {Koyama}, {Yamashita},
  {Morokuma-Matsui}, {Matsuhara}, {Nakagawa}, {Hayashi}, {Kodama}, {Shimakawa},
  {Suzuki}, {Tadaki}, {Tanaka}, \& {Yamamoto}}]{koyama2017}
{Koyama}, S., {Koyama}, Y., {Yamashita}, T., {et~al.} 2017, \apj, 847, 137

\bibitem[{{Koyama} {et~al.}(2014){Koyama}, {Kodama}, {Tadaki}, {Hayashi},
  {Tanaka}, \& {Shimakawa}}]{koyama2014}
{Koyama}, Y., {Kodama}, T., {Tadaki}, K.-i., {et~al.} 2014, \apj, 789, 18

\bibitem[{{Koyama} {et~al.}(2013){Koyama}, {Smail}, {Kurk}, {Geach}, {Sobral},
  {Kodama}, {Nakata}, {Swinbank}, {Best}, {Hayashi}, \& {Tadaki}}]{koyama2013}
{Koyama}, Y., {Smail}, I., {Kurk}, J., {et~al.} 2013, \mnras, 434, 423

\bibitem[{{Kulas} {et~al.}(2013){Kulas}, {McLean}, {Shapley}, {Steidel},
  {Konidaris}, {Matthews}, {Mace}, {Rudie}, {Trainor}, \& {Reddy}}]{kulas2013}
{Kulas}, K.~R., {McLean}, I.~S., {Shapley}, A.~E., {et~al.} 2013, \apj, 774,
  130

\bibitem[{{Laigle} {et~al.}(2016){Laigle}, {McCracken}, {Ilbert}, {Hsieh},
  {Davidzon}, {Capak}, {Hasinger}, {Silverman}, {Pichon}, {Coupon}, {Aussel},
  {Le Borgne}, {Caputi}, {Cassata}, {Chang}, {Civano}, {Dunlop}, {Fynbo},
  {Kartaltepe}, {Koekemoer}, {Le F{\`e}vre}, {Le Floc'h}, {Leauthaud}, {Lilly},
  {Lin}, {Marchesi}, {Milvang-Jensen}, {Salvato}, {Sanders}, {Scoville},
  {Smolcic}, {Stockmann}, {Taniguchi}, {Tasca}, {Toft}, {Vaccari}, \&
  {Zabl}}]{laigle2016}
{Laigle}, C., {McCracken}, H.~J., {Ilbert}, O., {et~al.} 2016, \apjs, 224, 24

\bibitem[{{Le F{\`e}vre} {et~al.}(2003){Le F{\`e}vre}, {Saisse}, {Mancini},
  {Brau-Nogue}, {Caputi}, {Castinel}, {D'Odorico}, {Garilli}, {Kissler-Patig},
  {Lucuix}, {Mancini}, {Pauget}, {Sciarretta}, {Scodeggio}, {Tresse}, \&
  {Vettolani}}]{lefevre2003}
{Le F{\`e}vre}, O., {Saisse}, M., {Mancini}, D., {et~al.} 2003, in \procspie,
  Vol. 4841, Instrument Design and Performance for Optical/Infrared
  Ground-based Telescopes, ed. M.~{Iye} \& A.~F.~M. {Moorwood}, 1670--1681

\bibitem[{{Lee} {et~al.}(2015){Lee}, {Im}, {Kim}, {Lotz}, {McPartland}, {Peth},
  \& {Koekemoer}}]{lee2015}
{Lee}, S.-K., {Im}, M., {Kim}, J.-W., {et~al.} 2015, \apj, 810, 90

\bibitem[{{Lemaux} {et~al.}(2014){Lemaux}, {Cucciati}, {Tasca}, {Le F{\`e}vre},
  {Zamorani}, {Cassata}, {Garilli}, {Le Brun}, {Maccagni}, {Pentericci},
  {Thomas}, {Vanzella}, {Zucca}, {Amor{\'{\i}}n}, {Bardelli}, {Capak},
  {Cassar{\`a}}, {Castellano}, {Cimatti}, {Cuby}, {de la Torre}, {Durkalec},
  {Fontana}, {Giavalisco}, {Grazian}, {Hathi}, {Ilbert}, {Moreau}, {Paltani},
  {Ribeiro}, {Salvato}, {Schaerer}, {Scodeggio}, {Sommariva}, {Talia},
  {Taniguchi}, {Tresse}, {Vergani}, {Wang}, {Charlot}, {Contini}, {Fotopoulou},
  {Gal}, {Kocevski}, {L{\'o}pez-Sanjuan}, {Lubin}, {Mellier}, {Sadibekova}, \&
  {Scoville}}]{lemaux2014}
{Lemaux}, B.~C., {Cucciati}, O., {Tasca}, L.~A.~M., {et~al.} 2014, \aap, 572,
  A41

\bibitem[{{Lewis} {et~al.}(2002){Lewis}, {Balogh}, {De Propris}, {Couch},
  {Bower}, {Offer}, {Bland-Hawthorn}, {Baldry}, {Baugh}, {Bridges}, {Cannon},
  {Cole}, {Colless}, {Collins}, {Cross}, {Dalton}, {Driver}, {Efstathiou},
  {Ellis}, {Frenk}, {Glazebrook}, {Hawkins}, {Jackson}, {Lahav}, {Lumsden},
  {Maddox}, {Madgwick}, {Norberg}, {Peacock}, {Percival}, {Peterson},
  {Sutherland}, \& {Taylor}}]{lewis2002}
{Lewis}, I., {Balogh}, M., {De Propris}, R., {et~al.} 2002, \mnras, 334, 673

\bibitem[{{Li} {et~al.}(2011){Li}, {Glazebrook}, {Gilbank}, {Balogh}, {Bower},
  {Baldry}, {Davies}, {Hau}, \& {McCarthy}}]{li2011}
{Li}, I.~H., {Glazebrook}, K., {Gilbank}, D., {et~al.} 2011, \mnras, 411, 1869

\bibitem[{{Li} {et~al.}(2009){Li}, {Yee}, \& {Ellingson}}]{li2009}
{Li}, I.~H., {Yee}, H.~K.~C., \& {Ellingson}, E. 2009, \apj, 698, 83

\bibitem[{{Lilly} {et~al.}(1996){Lilly}, {Le Fevre}, {Hammer}, \&
  {Crampton}}]{lilly1996}
{Lilly}, S.~J., {Le Fevre}, O., {Hammer}, F., \& {Crampton}, D. 1996, \apjl,
  460, L1

\bibitem[{{Lilly} {et~al.}(2007){Lilly}, {Le F{\`e}vre}, {Renzini}, {Zamorani},
  {Scodeggio}, {Contini}, {Carollo}, {Hasinger}, {Kneib}, {Iovino}, {Le Brun},
  {Maier}, {Mainieri}, {Mignoli}, {Silverman}, {Tasca}, {Bolzonella},
  {Bongiorno}, {Bottini}, {Capak}, {Caputi}, {Cimatti}, {Cucciati}, {Daddi},
  {Feldmann}, {Franzetti}, {Garilli}, {Guzzo}, {Ilbert}, {Kampczyk}, {Kovac},
  {Lamareille}, {Leauthaud}, {Borgne}, {McCracken}, {Marinoni}, {Pello},
  {Ricciardelli}, {Scarlata}, {Vergani}, {Sanders}, {Schinnerer}, {Scoville},
  {Taniguchi}, {Arnouts}, {Aussel}, {Bardelli}, {Brusa}, {Cappi}, {Ciliegi},
  {Finoguenov}, {Foucaud}, {Franceschini}, {Halliday}, {Impey}, {Knobel},
  {Koekemoer}, {Kurk}, {Maccagni}, {Maddox}, {Marano}, {Marconi}, {Meneux},
  {Mobasher}, {Moreau}, {Peacock}, {Porciani}, {Pozzetti}, {Scaramella},
  {Schiminovich}, {Shopbell}, {Smail}, {Thompson}, {Tresse}, {Vettolani},
  {Zanichelli}, \& {Zucca}}]{lilly2007}
{Lilly}, S.~J., {Le F{\`e}vre}, O., {Renzini}, A., {et~al.} 2007, \apjs, 172,
  70

\bibitem[{{Lubin} {et~al.}(2009){Lubin}, {Gal}, {Lemaux}, {Kocevski}, \&
  {Squires}}]{lubin2009}
{Lubin}, L.~M., {Gal}, R.~R., {Lemaux}, B.~C., {Kocevski}, D.~D., \& {Squires},
  G.~K. 2009, \aj, 137, 4867

\bibitem[{{Madau} \& {Dickinson}(2014)}]{madau2014}
{Madau}, P. \& {Dickinson}, M. 2014, \araa, 52, 415

\bibitem[{{Malavasi} {et~al.}(2017){Malavasi}, {Arnouts}, {Vibert}, {de la
  Torre}, {Moutard}, {Pichon}, {Davidzon}, {Kraljic}, {Bolzonella}, {Guzzo},
  {Garilli}, {Scodeggio}, {Granett}, {Abbas}, {Adami}, {Bottini}, {Cappi},
  {Cucciati}, {Franzetti}, {Fritz}, {Iovino}, {Krywult}, {Le Brun}, {Le
  F{\`e}vre}, {Maccagni}, {Ma{\l}ek}, {Marulli}, {Polletta}, {Pollo}, {Tasca},
  {Tojeiro}, {Vergani}, {Zanichelli}, {Bel}, {Branchini}, {Coupon}, {De Lucia},
  {Dubois}, {Hawken}, {Ilbert}, {Laigle}, {Moscardini}, {Sousbie}, {Treyer}, \&
  {Zamorani}}]{malavasi2017}
{Malavasi}, N., {Arnouts}, S., {Vibert}, D., {et~al.} 2017, \mnras, 465, 3817

\bibitem[{{Malavasi} {et~al.}(2016){Malavasi}, {Pozzetti}, {Cucciati},
  {Bardelli}, \& {Cimatti}}]{malavasi2016}
{Malavasi}, N., {Pozzetti}, L., {Cucciati}, O., {Bardelli}, S., \& {Cimatti},
  A. 2016, \aap, 585, A116

\bibitem[{{Masters} \& {Capak}(2011)}]{masters2011}
{Masters}, D. \& {Capak}, P. 2011, \pasp, 123, 638

\bibitem[{{Mihos} \& {Hernquist}(1996)}]{mihos1996}
{Mihos}, J.~C. \& {Hernquist}, L. 1996, \apj, 464, 641

\bibitem[{{Mok} {et~al.}(2013){Mok}, {Balogh}, {McGee}, {Wilman}, {Finoguenov},
  {Tanaka}, {Giodini}, {Bower}, {Connelly}, {Hou}, {Mulchaey}, \&
  {Parker}}]{mok2013}
{Mok}, A., {Balogh}, M.~L., {McGee}, S.~L., {et~al.} 2013, \mnras, 431, 1090

\bibitem[{{Mok} {et~al.}(2016){Mok}, {Wilson}, {Golding}, {Warren}, {Israel},
  {Serjeant}, {Knapen}, {S{\'a}nchez-Gallego}, {Barmby}, {Bendo}, {Rosolowsky},
  \& {van der Werf}}]{mok2016}
{Mok}, A., {Wilson}, C.~D., {Golding}, J., {et~al.} 2016, \mnras, 456, 4384

\bibitem[{{Moss}(2006)}]{moss2006}
{Moss}, C. 2006, \mnras, 373, 167

\bibitem[{{Mulroy} {et~al.}(2017){Mulroy}, {McGee}, {Gillman}, {Smith},
  {Haines}, {D{\'e}mocl{\`e}s}, {Okabe}, \& {Egami}}]{mulroy2017}
{Mulroy}, S.~L., {McGee}, S.~L., {Gillman}, S., {et~al.} 2017, \mnras, 472,
  3246

\bibitem[{{Muzzin} {et~al.}(2013){Muzzin}, {Marchesini}, {Stefanon}, {Franx},
  {Milvang-Jensen}, {Dunlop}, {Fynbo}, {Brammer}, {Labb{\'e}}, \& {van
  Dokkum}}]{muzzin2013}
{Muzzin}, A., {Marchesini}, D., {Stefanon}, M., {et~al.} 2013, \apjs, 206, 8

\bibitem[{{Muzzin} {et~al.}(2012){Muzzin}, {Wilson}, {Yee}, {Gilbank},
  {Hoekstra}, {Demarco}, {Balogh}, {van Dokkum}, {Franx}, {Ellingson}, {Hicks},
  {Nantais}, {Noble}, {Lacy}, {Lidman}, {Rettura}, {Surace}, \&
  {Webb}}]{muzzin2012}
{Muzzin}, A., {Wilson}, G., {Yee}, H.~K.~C., {et~al.} 2012, \apj, 746, 188

\bibitem[{{Nakata} {et~al.}(2005){Nakata}, {Bower}, {Balogh}, \&
  {Wilman}}]{nakata2005}
{Nakata}, F., {Bower}, R.~G., {Balogh}, M.~L., \& {Wilman}, D.~J. 2005, \mnras,
  357, 679

\bibitem[{{Noeske} {et~al.}(2007){Noeske}, {Weiner}, {Faber}, {Papovich},
  {Koo}, {Somerville}, {Bundy}, {Conselice}, {Newman}, {Schiminovich}, {Le
  Floc'h}, {Coil}, {Rieke}, {Lotz}, {Primack}, {Barmby}, {Cooper}, {Davis},
  {Ellis}, {Fazio}, {Guhathakurta}, {Huang}, {Kassin}, {Martin}, {Phillips},
  {Rich}, {Small}, {Willmer}, \& {Wilson}}]{noeske2007}
{Noeske}, K.~G., {Weiner}, B.~J., {Faber}, S.~M., {et~al.} 2007, \apjl, 660,
  L43

\bibitem[{{Oemler}(1974)}]{oemler1974}
{Oemler}, Jr., A. 1974, \apj, 194, 1

\bibitem[{{Oh} {et~al.}(2014){Oh}, {Mulchaey}, {Woo}, {Finoguenov}, {Tanaka},
  {Cooper}, {Ziparo}, {Bauer}, \& {Matsuoka}}]{oh2014}
{Oh}, S., {Mulchaey}, J.~S., {Woo}, J.-H., {et~al.} 2014, \apj, 790, 43

\bibitem[{{Owers} {et~al.}(2012){Owers}, {Couch}, {Nulsen}, \&
  {Randall}}]{owers2012}
{Owers}, M.~S., {Couch}, W.~J., {Nulsen}, P.~E.~J., \& {Randall}, S.~W. 2012,
  \apjl, 750, L23

\bibitem[{{Patel} {et~al.}(2009){Patel}, {Holden}, {Kelson}, {Illingworth}, \&
  {Franx}}]{patel2009}
{Patel}, S.~G., {Holden}, B.~P., {Kelson}, D.~D., {Illingworth}, G.~D., \&
  {Franx}, M. 2009, \apjl, 705, L67

\bibitem[{{Peng} {et~al.}(2010){Peng}, {Lilly}, {Kova{\v c}}, {Bolzonella},
  {Pozzetti}, {Renzini}, {Zamorani}, {Ilbert}, {Knobel}, {Iovino}, {Maier},
  {Cucciati}, {Tasca}, {Carollo}, {Silverman}, {Kampczyk}, {de Ravel},
  {Sanders}, {Scoville}, {Contini}, {Mainieri}, {Scodeggio}, {Kneib}, {Le
  F{\`e}vre}, {Bardelli}, {Bongiorno}, {Caputi}, {Coppa}, {de la Torre},
  {Franzetti}, {Garilli}, {Lamareille}, {Le Borgne}, {Le Brun}, {Mignoli},
  {Perez Montero}, {Pello}, {Ricciardelli}, {Tanaka}, {Tresse}, {Vergani},
  {Welikala}, {Zucca}, {Oesch}, {Abbas}, {Barnes}, {Bordoloi}, {Bottini},
  {Cappi}, {Cassata}, {Cimatti}, {Fumana}, {Hasinger}, {Koekemoer},
  {Leauthaud}, {Maccagni}, {Marinoni}, {McCracken}, {Memeo}, {Meneux}, {Nair},
  {Porciani}, {Presotto}, \& {Scaramella}}]{peng2010b}
{Peng}, Y.-j., {Lilly}, S.~J., {Kova{\v c}}, K., {et~al.} 2010, \apj, 721, 193

\bibitem[{{Pentericci} {et~al.}(2013){Pentericci}, {Castellano}, {Menci},
  {Salimbeni}, {Dahlen}, {Galametz}, {Santini}, {Grazian}, \&
  {Fontana}}]{pentericci2013}
{Pentericci}, L., {Castellano}, M., {Menci}, N., {et~al.} 2013, \aap, 552, A111

\bibitem[{{Perez} {et~al.}(2009){Perez}, {Tissera}, {Padilla}, {Alonso}, \&
  {Lambas}}]{perez2009}
{Perez}, J., {Tissera}, P., {Padilla}, N., {Alonso}, M.~S., \& {Lambas}, D.~G.
  2009, \mnras, 399, 1157

\bibitem[{{Poggianti} {et~al.}(2009){Poggianti}, {Arag{\'o}n-Salamanca},
  {Zaritsky}, {De Lucia}, {Milvang-Jensen}, {Desai}, {Jablonka}, {Halliday},
  {Rudnick}, {Varela}, {Bamford}, {Best}, {Clowe}, {Noll}, {Saglia},
  {Pell{\'o}}, {Simard}, {von der Linden}, \& {White}}]{poggianti2009}
{Poggianti}, B.~M., {Arag{\'o}n-Salamanca}, A., {Zaritsky}, D., {et~al.} 2009,
  \apj, 693, 112

\bibitem[{{Poggianti} {et~al.}(2006){Poggianti}, {von der Linden}, {De Lucia},
  {Desai}, {Simard}, {Halliday}, {Arag{\'o}n-Salamanca}, {Bower}, {Varela},
  {Best}, {Clowe}, {Dalcanton}, {Jablonka}, {Milvang-Jensen}, {Pello},
  {Rudnick}, {Saglia}, {White}, \& {Zaritsky}}]{poggianti2006}
{Poggianti}, B.~M., {von der Linden}, A., {De Lucia}, G., {et~al.} 2006, \apj,
  642, 188

\bibitem[{{Pradhan} {et~al.}(2006){Pradhan}, {Montenegro}, {Nahar}, \&
  {Eissner}}]{pradhan2006}
{Pradhan}, A.~K., {Montenegro}, M., {Nahar}, S.~N., \& {Eissner}, W. 2006,
  \mnras, 366, L6

\bibitem[{{Rodr{\'{\i}}guez del Pino} {et~al.}(2017){Rodr{\'{\i}}guez del
  Pino}, {Arag{\'o}n-Salamanca}, {Chies-Santos}, {Weinzirl}, {Bamford}, {Gray},
  {B{\"o}hm}, {Wolf}, \& {Maltby}}]{delpino2017}
{Rodr{\'{\i}}guez del Pino}, B., {Arag{\'o}n-Salamanca}, A., {Chies-Santos},
  A.~L., {et~al.} 2017, \mnras, 467, 4200

\bibitem[{{Roediger} {et~al.}(2014){Roediger}, {Br{\"u}ggen}, {Owers},
  {Ebeling}, \& {Sun}}]{roediger2014}
{Roediger}, E., {Br{\"u}ggen}, M., {Owers}, M.~S., {Ebeling}, H., \& {Sun}, M.
  2014, \mnras, 443, L114

\bibitem[{{Salpeter}(1955)}]{salpeter1955}
{Salpeter}, E.~E. 1955, \apj, 121, 161

\bibitem[{{Sanders} {et~al.}(2016){Sanders}, {Shapley}, {Kriek}, {Reddy},
  {Freeman}, {Coil}, {Siana}, {Mobasher}, {Shivaei}, {Price}, \& {de
  Groot}}]{sanders2016}
{Sanders}, R.~L., {Shapley}, A.~E., {Kriek}, M., {et~al.} 2016, \apj, 816, 23

\bibitem[{{Santos} {et~al.}(2013){Santos}, {Altieri}, {Popesso}, {Strazzullo},
  {Valtchanov}, {Berta}, {B{\"o}hringer}, {Conversi}, {Demarco}, {Edge},
  {Lidman}, {Lutz}, {Metcalfe}, {Mullis}, {Pintos-Castro},
  {S{\'a}nchez-Portal}, {Rawle}, {Rosati}, {Swinbank}, \&
  {Tanaka}}]{santos2013}
{Santos}, J.~S., {Altieri}, B., {Popesso}, P., {et~al.} 2013, \mnras, 433, 1287

\bibitem[{{Santos} {et~al.}(2014){Santos}, {Altieri}, {Tanaka}, {Valtchanov},
  {Saintonge}, {Dickinson}, {Foucaud}, {Kodama}, {Rawle}, \&
  {Tadaki}}]{santos2014}
{Santos}, J.~S., {Altieri}, B., {Tanaka}, M., {et~al.} 2014, \mnras, 438, 2565

\bibitem[{{Schaye} {et~al.}(2015){Schaye}, {Crain}, {Bower}, {Furlong},
  {Schaller}, {Theuns}, {Dalla Vecchia}, {Frenk}, {McCarthy}, {Helly},
  {Jenkins}, {Rosas-Guevara}, {White}, {Baes}, {Booth}, {Camps}, {Navarro},
  {Qu}, {Rahmati}, {Sawala}, {Thomas}, \& {Trayford}}]{schaye2015}
{Schaye}, J., {Crain}, R.~A., {Bower}, R.~G., {et~al.} 2015, \mnras, 446, 521

\bibitem[{{Scoville} {et~al.}(2013){Scoville}, {Arnouts}, {Aussel}, {Benson},
  {Bongiorno}, {Bundy}, {Calvo}, {Capak}, {Carollo}, {Civano}, {Dunlop},
  {Elvis}, {Faisst}, {Finoguenov}, {Fu}, {Giavalisco}, {Guo}, {Ilbert},
  {Iovino}, {Kajisawa}, {Kartaltepe}, {Leauthaud}, {Le F{\`e}vre}, {LeFloch},
  {Lilly}, {Liu}, {Manohar}, {Massey}, {Masters}, {McCracken}, {Mobasher},
  {Peng}, {Renzini}, {Rhodes}, {Salvato}, {Sanders}, {Sarvestani}, {Scarlata},
  {Schinnerer}, {Sheth}, {Shopbell}, {Smol{\v c}i{\'c}}, {Taniguchi}, {Taylor},
  {White}, \& {Yan}}]{scoville2013}
{Scoville}, N., {Arnouts}, S., {Aussel}, H., {et~al.} 2013, \apjs, 206, 3

\bibitem[{{Seaton} \& {Osterbrock}(1957)}]{seaton1957}
{Seaton}, M.~J. \& {Osterbrock}, D.~E. 1957, \apj, 125, 66

\bibitem[{{Serra} {et~al.}(2012){Serra}, {Oosterloo}, {Morganti}, {Alatalo},
  {Blitz}, {Bois}, {Bournaud}, {Bureau}, {Cappellari}, {Crocker}, {Davies},
  {Davis}, {de Zeeuw}, {Duc}, {Emsellem}, {Khochfar}, {Krajnovi{\'c}},
  {Kuntschner}, {Lablanche}, {McDermid}, {Naab}, {Sarzi}, {Scott}, {Trager},
  {Weijmans}, \& {Young}}]{serra2012}
{Serra}, P., {Oosterloo}, T., {Morganti}, R., {et~al.} 2012, \mnras, 422, 1835

\bibitem[{{Shimakawa} {et~al.}(2018){Shimakawa}, {Kodama}, {Hayashi},
  {Prochaska}, {Tanaka}, {Cai}, {Suzuki}, {Tadaki}, \&
  {Koyama}}]{shimakawa2018}
{Shimakawa}, R., {Kodama}, T., {Hayashi}, M., {et~al.} 2018, \mnras, 473, 1977

\bibitem[{{Shimakawa} {et~al.}(2015){Shimakawa}, {Kodama}, {Tadaki}, {Hayashi},
  {Koyama}, \& {Tanaka}}]{shimakawa2015}
{Shimakawa}, R., {Kodama}, T., {Tadaki}, K.-i., {et~al.} 2015, \mnras, 448, 666

\bibitem[{{Sobral} {et~al.}(2009){Sobral}, {Best}, {Geach}, {Smail}, {Kurk},
  {Cirasuolo}, {Casali}, {Ivison}, {Coppin}, \& {Dalton}}]{sobral2009}
{Sobral}, D., {Best}, P.~N., {Geach}, J.~E., {et~al.} 2009, \mnras, 398, 75

\bibitem[{{Sobral} {et~al.}(2011){Sobral}, {Best}, {Smail}, {Geach},
  {Cirasuolo}, {Garn}, \& {Dalton}}]{sobral2011}
{Sobral}, D., {Best}, P.~N., {Smail}, I., {et~al.} 2011, \mnras, 411, 675

\bibitem[{{Sobral} {et~al.}(2013){Sobral}, {Smail}, {Best}, {Geach}, {Matsuda},
  {Stott}, {Cirasuolo}, \& {Kurk}}]{sobral2013}
{Sobral}, D., {Smail}, I., {Best}, P.~N., {et~al.} 2013, \mnras, 428, 1128

\bibitem[{{Sobral} {et~al.}(2015){Sobral}, {Stroe}, {Dawson}, {Wittman}, {Jee},
  {R{\"o}ttgering}, {van Weeren}, \& {Br{\"u}ggen}}]{sobral2015b}
{Sobral}, D., {Stroe}, A., {Dawson}, W.~A., {et~al.} 2015, \mnras, 450, 630

\bibitem[{{Sobral} {et~al.}(2016){Sobral}, {Stroe}, {Koyama}, {Darvish},
  {Calhau}, {Afonso}, {Kodama}, \& {Nakata}}]{sobral2016}
{Sobral}, D., {Stroe}, A., {Koyama}, Y., {et~al.} 2016, \mnras, 458, 3443

\bibitem[{{Stroe} \& {Sobral}(2015)}]{stroe2015}
{Stroe}, A. \& {Sobral}, D. 2015, \mnras, 453, 242

\bibitem[{{Stroe} {et~al.}(2015){Stroe}, {Sobral}, {Dawson}, {Jee}, {Hoekstra},
  {Wittman}, {van Weeren}, {Br{\"u}ggen}, \& {R{\"o}ttgering}}]{stroe2015a}
{Stroe}, A., {Sobral}, D., {Dawson}, W., {et~al.} 2015, \mnras, 450, 646

\bibitem[{{Stroe} {et~al.}(2017){Stroe}, {Sobral}, {Paulino-Afonso}, {Alegre},
  {Calhau}, {Santos}, \& {van Weeren}}]{stroe2017}
{Stroe}, A., {Sobral}, D., {Paulino-Afonso}, A., {et~al.} 2017, \mnras, 465,
  2916

\bibitem[{{Stroe} {et~al.}(2014){Stroe}, {Sobral}, {R{\"o}ttgering}, \& {van
  Weeren}}]{stroe2014}
{Stroe}, A., {Sobral}, D., {R{\"o}ttgering}, H.~J.~A., \& {van Weeren}, R.~J.
  2014, \mnras, 438, 1377

\bibitem[{{Tadaki} {et~al.}(2012){Tadaki}, {Kodama}, {Ota}, {Hayashi},
  {Koyama}, {Papovich}, {Brodwin}, {Tanaka}, \& {Iye}}]{tadaki2012}
{Tadaki}, K.-i., {Kodama}, T., {Ota}, K., {et~al.} 2012, \mnras, 423, 2617

\bibitem[{{Tonnesen} \& {Cen}(2012)}]{tonnesen2012}
{Tonnesen}, S. \& {Cen}, R. 2012, \mnras, 425, 2313

\bibitem[{{Tran} {et~al.}(2010){Tran}, {Papovich}, {Saintonge}, {Brodwin},
  {Dunlop}, {Farrah}, {Finkelstein}, {Finkelstein}, {Lotz}, {McLure},
  {Momcheva}, \& {Willmer}}]{tran2010}
{Tran}, K.-V.~H., {Papovich}, C., {Saintonge}, A., {et~al.} 2010, \apjl, 719,
  L126

\bibitem[{{Treu} {et~al.}(2003){Treu}, {Ellis}, {Kneib}, {Dressler}, {Smail},
  {Czoske}, {Oemler}, \& {Natarajan}}]{treu2003}
{Treu}, T., {Ellis}, R.~S., {Kneib}, J.-P., {et~al.} 2003, \apj, 591, 53

\bibitem[{{van der Wel} {et~al.}(2016){van der Wel}, {Noeske}, {Bezanson},
  {Pacifici}, {Gallazzi}, {Franx}, {Mu{\~n}oz-Mateos}, {Bell}, {Brammer},
  {Charlot}, {Chauk{\'e}}, {Labb{\'e}}, {Maseda}, {Muzzin}, {Rix}, {Sobral},
  {van de Sande}, {van Dokkum}, {Wild}, \& {Wolf}}]{vanderwel2016}
{van der Wel}, A., {Noeske}, K., {Bezanson}, R., {et~al.} 2016, \apjs, 223, 29

\bibitem[{{Vogelsberger} {et~al.}(2014){Vogelsberger}, {Genel}, {Springel},
  {Torrey}, {Sijacki}, {Xu}, {Snyder}, {Bird}, {Nelson}, \&
  {Hernquist}}]{volgersberger2014}
{Vogelsberger}, M., {Genel}, S., {Springel}, V., {et~al.} 2014, \nat, 509, 177

\bibitem[{Walt {et~al.}(2011)Walt, Colbert, \& Varoquaux}]{walt2011}
Walt, S. v.~d., Colbert, S.~C., \& Varoquaux, G. 2011, Computing in Science \&
  Engineering, 13, 22

\bibitem[{{Whitaker} {et~al.}(2012){Whitaker}, {van Dokkum}, {Brammer}, \&
  {Franx}}]{whitaker2012}
{Whitaker}, K.~E., {van Dokkum}, P.~G., {Brammer}, G., \& {Franx}, M. 2012,
  \apjl, 754, L29

\bibitem[{{White} {et~al.}(2005){White}, {Clowe}, {Simard}, {Rudnick}, {De
  Lucia}, {Arag{\'o}n-Salamanca}, {Bender}, {Best}, {Bremer}, {Charlot},
  {Dalcanton}, {Dantel}, {Desai}, {Fort}, {Halliday}, {Jablonka}, {Kauffmann},
  {Mellier}, {Milvang-Jensen}, {Pell{\'o}}, {Poggianti}, {Poirier},
  {Rottgering}, {Saglia}, {Schneider}, \& {Zaritsky}}]{white2005}
{White}, S.~D.~M., {Clowe}, D.~I., {Simard}, L., {et~al.} 2005, \aap, 444, 365

\bibitem[{{Wijesinghe} {et~al.}(2012){Wijesinghe}, {Hopkins}, {Brough},
  {Taylor}, {Norberg}, {Bauer}, {Brown}, {Cameron}, {Conselice}, {Croom},
  {Driver}, {Grootes}, {Jones}, {Kelvin}, {Loveday}, {Pimbblet}, {Popescu},
  {Prescott}, {Sharp}, {Baldry}, {Sadler}, {Liske}, {Robotham}, {Bamford},
  {Bland-Hawthorn}, {Gunawardhana}, {Meyer}, {Parkinson}, {Drinkwater},
  {Peacock}, \& {Tuffs}}]{wijesinghe2012}
{Wijesinghe}, D.~B., {Hopkins}, A.~M., {Brough}, S., {et~al.} 2012, \mnras,
  423, 3679

\bibitem[{{Wolf} {et~al.}(2009){Wolf}, {Arag{\'o}n-Salamanca}, {Balogh},
  {Barden}, {Bell}, {Gray}, {Peng}, {Bacon}, {Barazza}, {B{\"o}hm}, {Caldwell},
  {Gallazzi}, {H{\"a}u{\ss}ler}, {Heymans}, {Jahnke}, {Jogee}, {van Kampen},
  {Lane}, {McIntosh}, {Meisenheimer}, {Papovich}, {S{\'a}nchez}, {Taylor},
  {Wisotzki}, \& {Zheng}}]{wolf2009}
{Wolf}, C., {Arag{\'o}n-Salamanca}, A., {Balogh}, M., {et~al.} 2009, \mnras,
  393, 1302

\bibitem[{{Yan} {et~al.}(2006){Yan}, {Newman}, {Faber}, {Konidaris}, {Koo}, \&
  {Davis}}]{yan2006}
{Yan}, R., {Newman}, J.~A., {Faber}, S.~M., {et~al.} 2006, \apj, 648, 281

\end{thebibliography}


\begin{appendix}

\section{Sample Completeness}\label{sec:completeness}

\begin{figure}
\centering
\includegraphics[width=\linewidth]{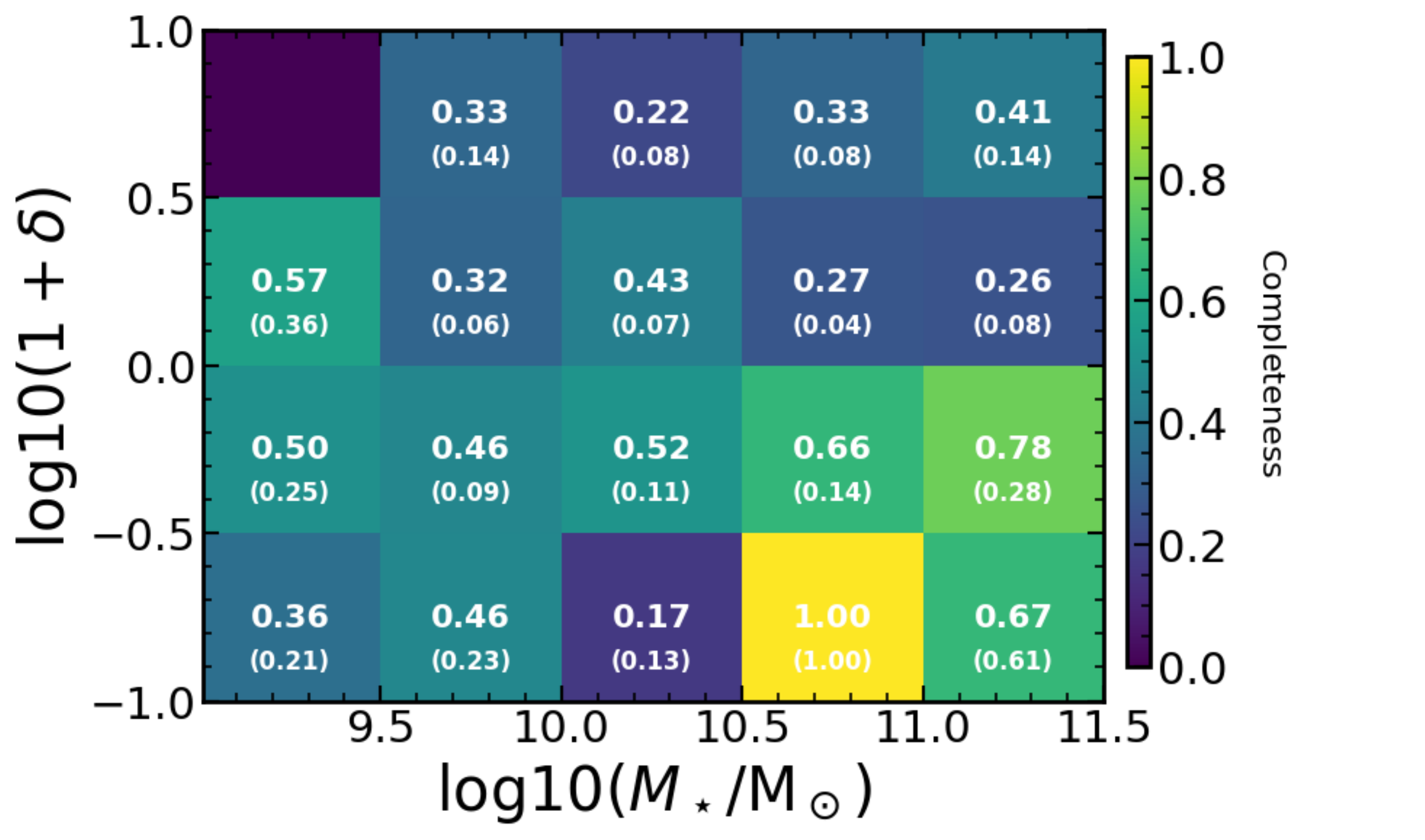}
\includegraphics[width=\linewidth]{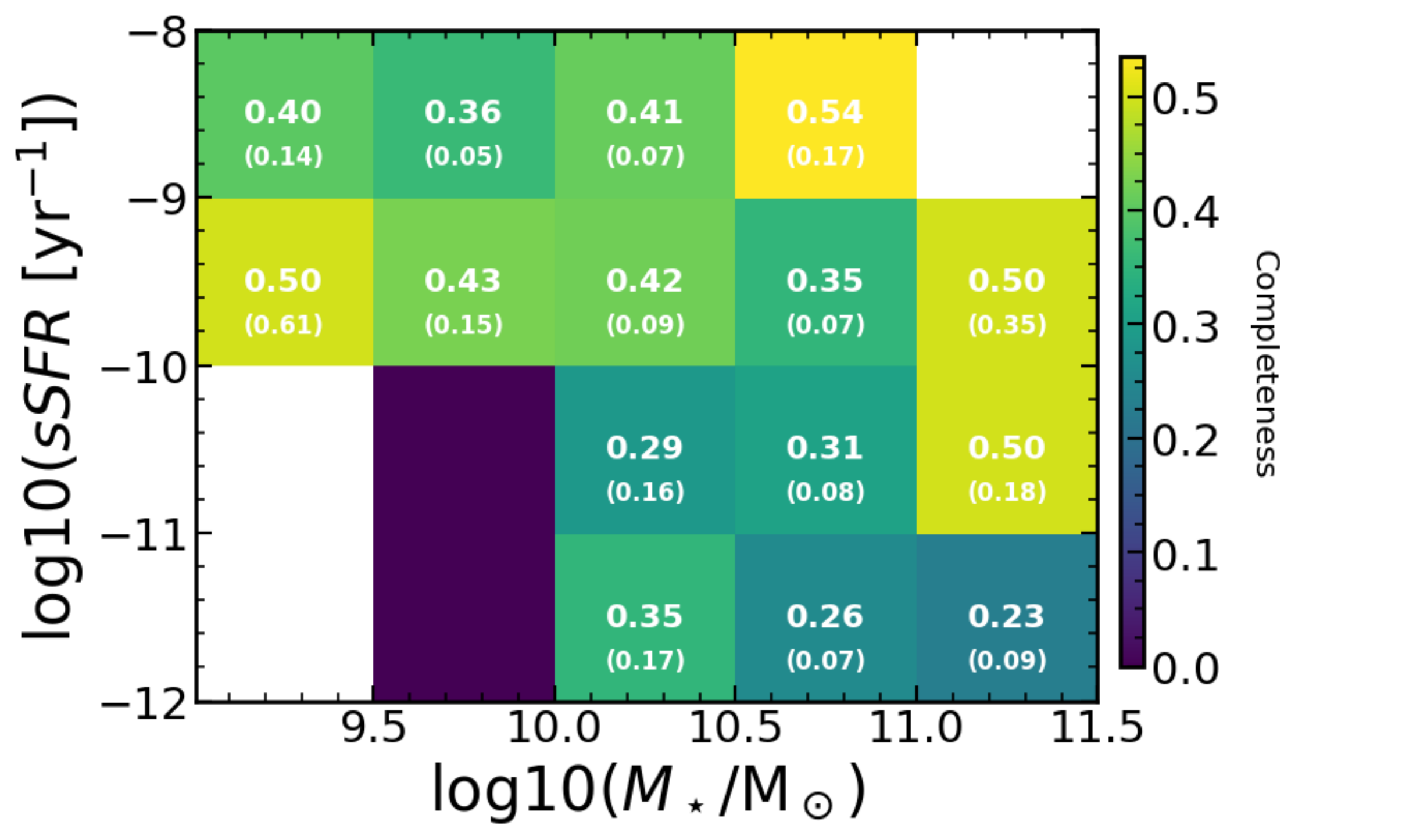}
\caption{Spectroscopic sample completeness as a function of stellar mass and local overdensity (top) and specific star formation rate (bottom). In each panel we indicate in white numbers the completeness for each bin (Poisson errors shown in parenthesis). Bins with no targets are shown in white.}
\label{fig:sample_completeness}
\end{figure}

We estimate the sample completeness of our spectroscopic observations by comparing the number of sources for which we successfully measured a redshift with the number of possible targets in the parent catalogue (given our selection described in Section \ref{ssection:selection}). We present our results in Figure \ref{fig:sample_completeness}. We will discuss the completeness effects in more detail in a forthcoming paper.

We confirm that we are under-sampling denser regions when compared to the lowest density regions, which is expected given the spatial constraints on the positioning of the slits in the VIMOS masks does not allow to target densely populated areas. In terms of star formation activity we find that our typical completeness is lower for quiescent galaxies ($\sim30\%$) when compared to star-forming ones ($\sim40\%$). When taken together we find that we are most likely missing quiescent galaxies in high density regions, but that the difference between the two populations is not dramatic in terms of completeness and our derived completeness corrections can tackle this without problems. Therefore we are providing a fair representation of the galaxy population in the regions we are targeting.

\section{Catalogue of superstructure members}

We release with this paper the VIS$^3$COS catalogue of all targets in and around the superstructure at $z\sim1$ with spectroscopic redshifts, along with some of their measured properties: SFR, overdensity, stellar mass. We present the first 10 entries of the full catalogue in Table \ref{table:Gal_props}.

\begin{table*}
\caption{First 10 galaxies in our sample. The first column is our catalogue ID. The second and third two columns show the object coordinates from \citet{ilbert2009}. The fourth column is our measured spectroscopic redshift. The fifth column is the $K$-band magnitude from \citet{ilbert2009}. The sixth and seventh columns are the stellar masses and SFRs derived with \textsc{magphys}. The last column is the local overdensity from \citet{darvish2015a,darvish2017}.}
\label{table:Gal_props}
\begin{center}
\begin{tabular}{lcccccccc}
\hline
ID & RA & DEC & $z_\mathrm{spec}$ & $K_\mathrm{AB}$ & $\log_{10}(M_\star)$ & $\log_{10}(SFR)$ & $\log_{10}(1+\delta)$ \\
 & (J2000) & (J2000) &  &  & $(M_\odot)$ & $(M_\odot yr^{-1})$ &  \\
\hline
VIS3COS-1 & 150.510640 & 2.035566 & 0.8998 & 20.6 & 10.6 & 0.9 & -0.26 \\
VIS3COS-2 & 150.521776 & 2.040788 & 0.8707 & 21.0 & 10.4 & 0.1 & 0.15 \\
VIS3COS-3 & 150.547778 & 2.044605 & 0.8714 & 20.5 & 11.0 & 0.5 & 0.39 \\
VIS3COS-4 & 150.543696 & 2.047819 & 0.8080 & 21.9 & 9.2 & 0.6 & -0.68 \\
VIS3COS-6 & 150.590194 & 2.051661 & 0.8419 & 20.0 & 10.6 & 1.1 & 0.11 \\
VIS3COS-7 & 150.573302 & 2.053824 & 0.8722 & 21.5 & 10.4 & 1.3 & -0.19 \\
VIS3COS-8 & 150.520207 & 2.057174 & 0.8970 & 21.1 & 10.1 & 1.1 & 0.28 \\
VIS3COS-10 & 150.569219 & 2.062002 & 0.6980 & 19.7 & 11.2 & 1.5 & -0.54 \\
VIS3COS-12 & 150.575212 & 2.068181 & 0.8724 & 99.0 & 10.5 & -0.7 & -0.09 \\
VIS3COS-13 & 150.538943 & 2.070524 & 0.8930 & 20.5 & 10.8 & 0.2 & 0.14 \\
VIS3COS-15 & 150.604440 & 2.074035 & 0.8555 & 21.9 & 9.6 & 0.3 & 0.03 \\
\hline
\end{tabular}
\end{center}
\end{table*}

\section{Individual Stacks}

\begin{figure}
\centering
\includegraphics[width=\linewidth]{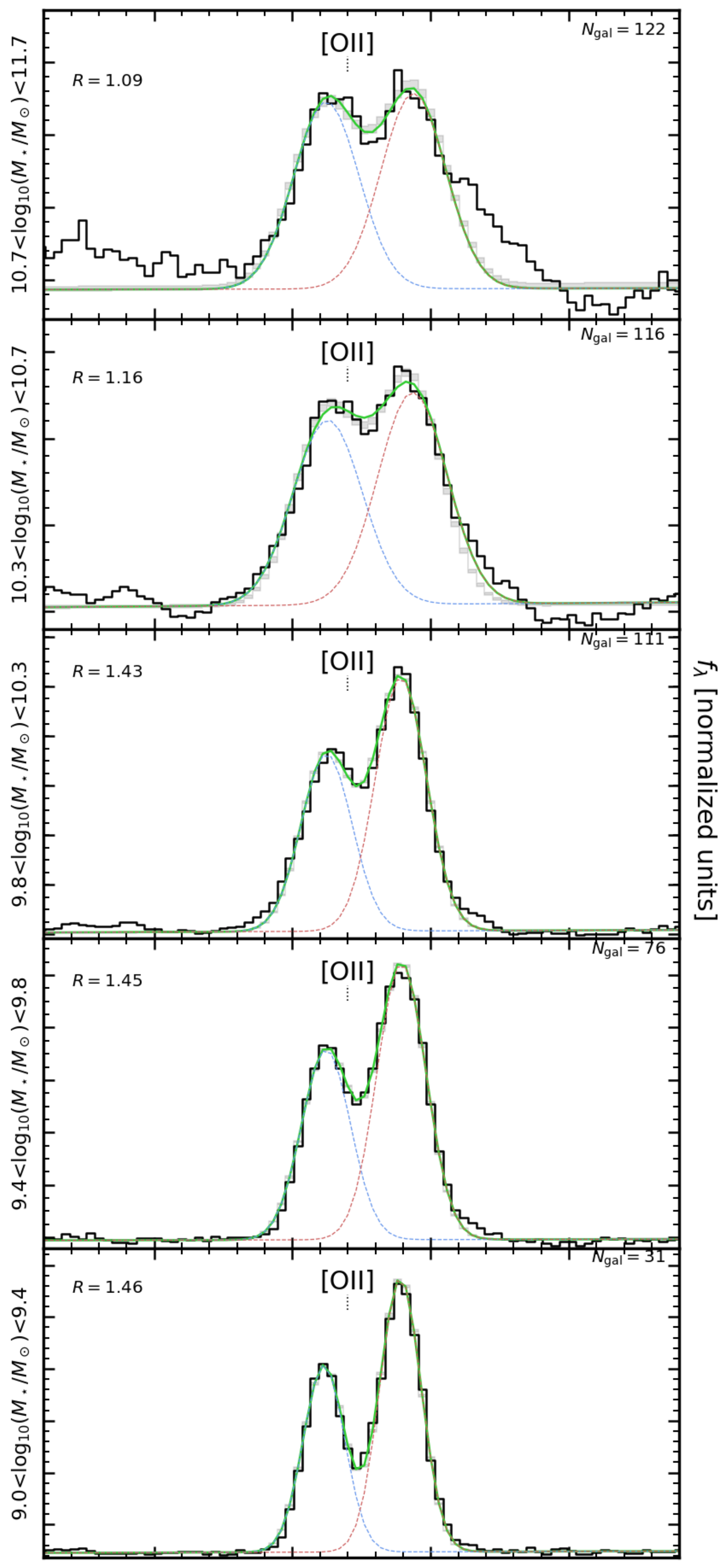}
\caption{Resulting spectral stacks (solid black line) in bins of stellar mass (high to low stellar mass from top to bottom) around the [O{\sc ii}] doublet. We show in green the best fit doublet model with each component shown as blue and red dashed lines. The shaded grey area represents the typical error on the fit of the spectra at each wavelength computed from the 16th and 84th percentiles of 10000 realizations of perturbing the spectra by its error. In each panel we show the derived ratio between the two doublet components.}
\label{fig:stack_Mass_full}
\end{figure}

\begin{figure}
\centering
\includegraphics[width=\linewidth]{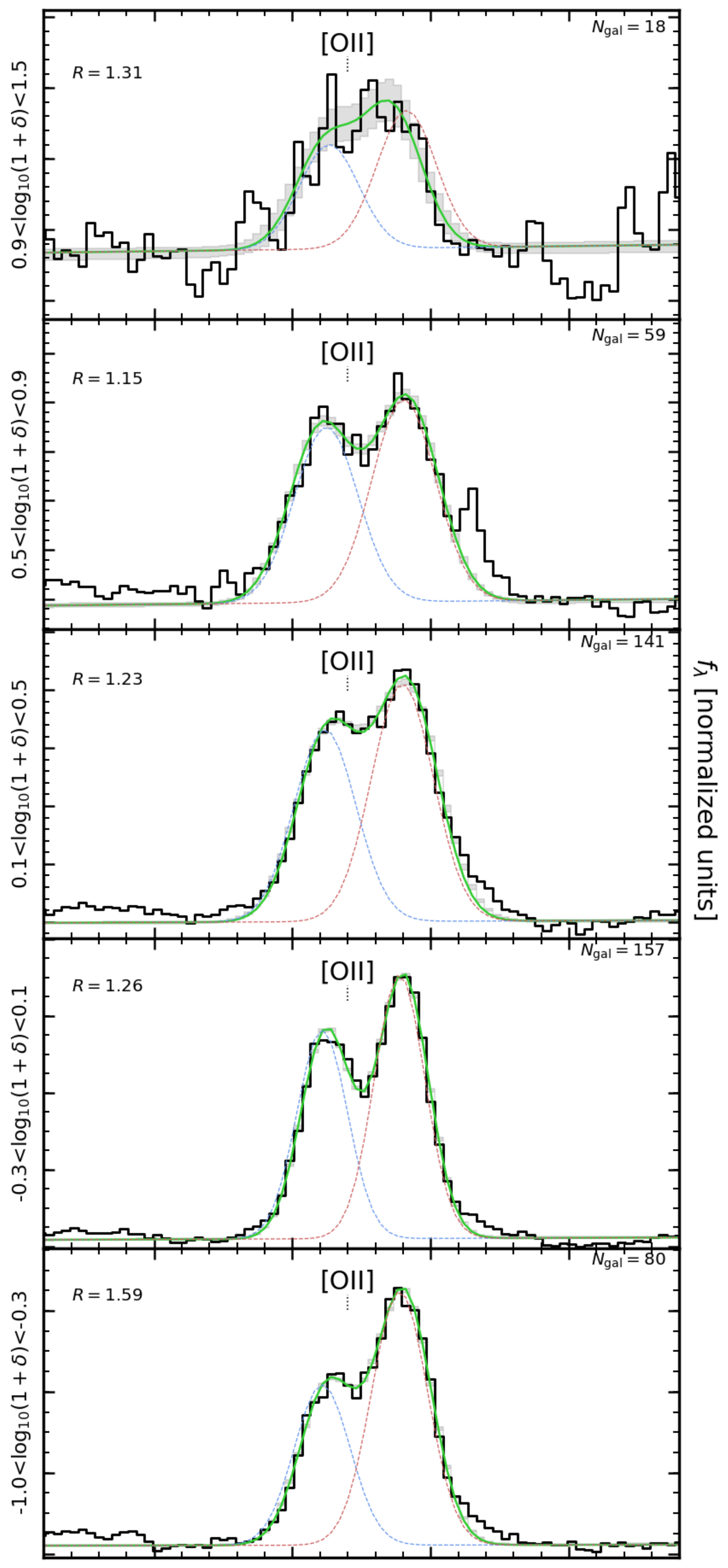}
\caption{Resulting spectral stacks (solid black line) in bins of overdensity (high to low density from top to bottom) around the [O{\sc ii}] doublet. We show in green the best fit doublet model with each component shown as blue and red dashed lines. The shaded grey area represents the typical error on the fit of the spectra at each wavelength computed from the 16th and 84th percentiles of 10000 realizations of perturbing the spectra by its error. In each panel we show the derived ratio between the two doublet components.}
\label{fig:stack_delta_full}
\end{figure}

Since some trends are difficult to see when showing all stacked spectra in a single panel due to line cluttering, we show in this Appendix all the stacked spectra individually in Figures \ref{fig:stack_Mass_full} (in bins of stellar mass) and \ref{fig:stack_delta_full} (in bins of overdensity). All results are also summarized in Table \ref{table:line_props}.

\begin{table*}
\begin{center}
\caption{Summary of [O{\sc ii}] properties from the stacked spectra. Equivalent widths (EW) are in units of \AA. The third column shows the doublet ratio $R = $[O{\sc ii}]$\lambda3729/\lambda3726$.}
\begin{tabular}{ccc}
\hline
Range & EW([O{\sc ii}])   & $R$ \\
\hline
$9.0 < \log_{10}\left(M_\star/M_\odot\right) < 9.4 $& $ -35.4_{-0.4}^{+0.4} $&$ 1.46_{-0.07}^{+0.07} $ \\
$9.4 < \log_{10}\left(M_\star/M_\odot\right) < 9.8 $& $ -20.7_{-0.2}^{+0.2} $&$ 1.45_{-0.06}^{+0.06} $ \\
$9.8 < \log_{10}\left(M_\star/M_\odot\right) < 10.3 $& $ -18.9_{-0.2}^{+0.2} $&$ 1.43_{-0.06}^{+0.06} $ \\
$10.3 < \log_{10}\left(M_\star/M_\odot\right) < 10.7 $& $ -6.9_{-0.2}^{+0.3} $&$ 1.16_{-0.10}^{+0.10} $ \\
$10.7 < \log_{10}\left(M_\star/M_\odot\right) < 11.7 $& $ -4.0_{-0.2}^{+0.2} $&$ 1.09_{-0.08}^{+0.08} $ \\
\hline
$-1.0 < \log_{10}(1+\delta) < -0.3 $& $ -11.7_{-0.2}^{+0.2} $&$ 1.59_{-0.07}^{+0.07} $ \\
$-0.3 < \log_{10}(1+\delta) < 0.1 $& $ -12.8_{-0.2}^{+0.2} $&$ 1.26_{-0.05}^{+0.05} $ \\
$0.1 < \log_{10}(1+\delta) < 0.5 $& $ -10.4_{-0.4}^{+0.4} $&$ 1.23_{-0.08}^{+0.08} $ \\
$0.5 < \log_{10}(1+\delta) < 0.9 $& $ -5.6_{-0.3}^{+0.3} $&$ 1.15_{-0.09}^{+0.09} $ \\
$0.9 < \log_{10}(1+\delta) < 1.5 $& $ -2.5_{-0.8}^{+1.1} $&$ 1.31_{-0.34}^{+0.47} $ \\
\hline
\end{tabular}
\label{table:line_props}
\end{center}
\end{table*}

\end{appendix}
\end{document}